\documentclass[a4paper,12pt]{article}
\pdfoutput=1 
\usepackage{jheppub} 
\usepackage{epsf}
\usepackage{amsmath,amssymb}
\input{colordvi.tex}
\usepackage[usenames,dvipsnames]{color}
\usepackage{graphicx}
\usepackage{ascmac}
\usepackage{hyperref}
\usepackage{slashed}
\usepackage{tikz-feynman}
\tikzfeynmanset{compat=1.0.0}

\begin{document}

\begin{flushright}
TU-1306,
KEK-QUP-2026-0015
\end{flushright}

\title{Gravitational Waves from New Inflation}

\author[a,b]{Naoya Kitajima,}
\author[a]{Kamil Mudrunka,}
\author[a,c]{Kazunori Nakayama}

\affiliation[a]{Department of Physics, Tohoku University, Sendai 980-8578, Japan}
\affiliation[b]{Frontier Research Institute for Interdisciplinary Sciences, Tohoku University,
6-3 Azaaoba, Aramaki, Aoba-ku, Sendai 980-8578, Japan}
\affiliation[c]{International Center for Quantum-field Measurement Systems for Studies of the Universe and Particles (QUP), KEK, 1-1 Oho, Tsukuba, Ibaraki 305-0801, Japan}

\emailAdd{naoya.kitajima.c2@tohoku.ac.jp}
\emailAdd{mudrunka.kamil.q5@dc.tohoku.ac.jp}
\emailAdd{kazunori.nakayama.d3@tohoku.ac.jp}

\abstract{Gravitational waves provide one of the most powerful observational probes for discriminating among inflationary models. In particular, it has recently become clear that the high-frequency gravitational-wave spectrum encodes a wealth of information about the properties of the inflaton and the post-inflationary dynamics. In this paper, we present the first comprehensive prediction of the complete gravitational-wave spectrum in the new inflation scenario, a prototypical realization of small-field inflation, incorporating all known production mechanisms including both the classical and quantum sources.}

\maketitle
\flushbottom

\section{Introduction}

Among the numerous inflationary models proposed over the past decades~\cite{Martin:2013tda}, new inflation~\cite{Linde:1981mu,Albrecht:1982wi}, or equivalently hilltop inflation, is a prototypical realization of small-field inflation and remains one of the phenomenologically viable scenarios. It naturally predicts a scalar spectral index consistent with the latest observations of the cosmic microwave background (CMB)~\cite{Planck:2018vyg}, while its inflationary energy scale is typically low~\cite{Kumekawa:1994gx,Izawa:1996dv,Asaka:1999jb,Senoguz:2004ky,Kohri:2007gq,Nakayama:2012dw,Ema:2017rkk,Dimopoulos:2020kol}. Consequently, the primordial tensor-to-scalar ratio is generally too small to be detectable by present or future CMB experiments.

In contrast, the post-inflationary reheating dynamics can give rise to a variety of nontrivial phenomena that produce high-frequency gravitational waves (GWs). Unlike the primordial tensor perturbations generated during inflation~\cite{Starobinsky:1979ty}, the amplitudes of these GWs are not directly tied to the inflationary energy scale and can therefore reach observable levels even in low-scale inflation models. The primary goal of this work is to provide a comprehensive prediction of the gravitational-wave signals expected from the new inflation scenario.

One of the most extensively studied sources of high-frequency GWs is preheating~\cite{Kofman:1994rk,Kofman:1997yn}. Following the end of inflation, the inflaton field can undergo explosive particle production, leading to the rapid growth of inhomogeneous field fluctuations that act as classical sources of GWs~\cite{Khlebnikov:1997di,Easther:2006vd,Easther:2006gt,Garcia-Bellido:2007nns,Garcia-Bellido:2007fiu,Dufaux:2007pt,Dufaux:2008dn}. In many realizations of new inflation, this process proceeds through efficient tachyonic preheating~\cite{Felder:2001kt,Felder:2000hj,Desroche:2005yt,Brax:2010ai,Antusch:2015nla,Antusch:2015vna}. We investigate the resulting GW production by performing classical lattice simulations of the preheating dynamics.

Recently, increasing attention has been devoted to high-frequency GWs generated through quantum processes during reheating. Several production mechanisms have been identified, including graviton emission from inflaton annihilation~\cite{Ema:2015dka,Ema:2016hlw,Ema:2020ggo,Choi:2024ilx,Mudrunka:2026kgm,Wang:2026ule,Wang:2026pff}, bremsstrahlung accompanying inflaton decay~\cite{Nakayama:2018ptw,Huang:2019lgd,Barman:2023ymn,Barman:2023rpg,Hu:2024awd,Hu:2024bha,Lee:2026fpg}, graviton production via scatterings between inflatons and Standard Model (SM) particles in the thermal plasma~\cite{Xu:2024fjl,Bernal:2025lxp,Xu:2025wjq}, and scatterings among SM particles themselves also produce gravitons~\cite{Ghiglieri:2015nfa,Ghiglieri:2020mhm,Ringwald:2020ist,Ghiglieri:2022rfp}. These mechanisms are expected to generate GWs at frequencies substantially higher than those associated with preheating.

In this paper, we present a comprehensive analysis of all of these GW production mechanisms within the framework of the new inflation model. By computing the corresponding GW spectra over a wide range of frequencies, we derive robust and characteristic predictions that can serve as distinctive signatures of the new inflation scenario, thereby providing the first unified prediction spanning both the classical and quantum regimes.

This paper is organized as follows.
We briefly review the basics of new inflation in Sec.~\ref{sec:new} and GW spectrum in Sec.~\ref{sec:gw_def}.
In Sec.~\ref{sec:gw_p} we calculate the nonlinear dynamics of the inflaton after new inflation and resulting GW spectrum with detailed lattice simulations.
In Sec.~\ref{sec:gw_q} we compute GW spectra from quantum origins, including primordial GWs in both low and high frequency regime, bremsstrahlung GWs and GWs from particle scatterings.
In Sec.~\ref{sec:full} we combine all the results to show full GW spectra predicted in the new inflation model.
Sec.~\ref{sec:conc} is devoted to conclusions.

\section{New inflation}\label{sec:new}

We define the new inflation model by the following potential~\cite{Kumekawa:1994gx,Izawa:1996dv,Asaka:1999jb,Senoguz:2004ky}:
\begin{align}
	V = \Lambda^4 \left[ 1 - \left(\frac{\phi}{v}\right)^n \right]^2,
    \label{V}
\end{align}
where $\phi$ is canonically normalized real scalar inflaton field, $v$ and $\Lambda$ are constants and $n$ is an integer. This form of the potential is motivated by supersymmetry~\cite{Kumekawa:1994gx,Izawa:1996dv,Asaka:1999jb,Senoguz:2004ky,Nakayama:2012dw}.
Inflation happens at small field value $\phi\ll v$ (without loss of generality we can take $\phi>0$), while slowly rolling toward the minimum of the potential $\phi=v$.\footnote{
    Such an initial condition of the inflaton may be explained by the so-called preinflation~\cite{Izawa:1997df}.}
The inflationary dynamics is analyzed in a standard way using slow-roll parameters $\epsilon \equiv \frac{M_{\rm Pl}^2}{2}\left(\frac{V'}{V}\right)^2$ and $\eta \equiv M_{\rm Pl}^2\frac{V''}{V}$~\cite{Liddle:2000cg} with $M_{\rm Pl}$ being the reduced Planck scale.
The field value at the end of inflation, $\phi_{\rm e}$, is defined by the condition $|\eta|=1$ and given by 
\begin{align}
	\phi_{\rm e} = v \left[ \frac{1}{2n(n-1)} \frac{v^2}{M_{\rm Pl}^2} \right]^{1/(n-2)}.
\end{align}
The field value at the e-folding number $N$ before the end of inflation is given by
\begin{align}
	\phi (N) = v \left[ \frac{1}{2n(n-2) N} \frac{v^2}{M_{\rm Pl}^2} \right]^{1/(n-2)}.
\end{align}
The generation of long wavelength scalar and tensor fluctuations is also calculated in a standard way.  
The dimensionless power spectrum of the curvature perturbation is given by
\begin{align}
	P_\zeta = \frac{1}{12\pi^2}\left[2n \left((n-2) N\right)^{n-1}\right]^{\frac{2}{n-2}} \frac{\Lambda^4}{(v^n M_{\rm Pl}^{n-4})^{\frac{2}{n-2}}}.
\end{align}
The CMB normalization gives $P_\zeta \simeq 2.2\times 10^{-9}$~\cite{Planck:2018vyg}.
The scalar spectral index and the tensor-to-scalar ratio are calculated as
\begin{align}
	n_s = 1- \frac{2}{N}\frac{n-1}{n-2},~~~~~~r =\frac{16n}{N(n-2)} \left[\frac{1}{2n(n-2)N}\frac{v^2}{M_{\rm Pl}^2}\right]^{\frac{n}{n-2}}.
\end{align}
In order for $n_s$ to be consistent with the CMB observation, $n\geq 6$ may be required in this minimal setup. However, one can introduce a small modification term to the potential so that $n_s$ fits with the observation even for $n=4$~\cite{Nakayama:2012dw,Takahashi:2013cxa}.\footnote{
    These analytical results assume $v\lesssim M_{\rm Pl}$. If $v\gtrsim M_{\rm Pl}$ the prediction changes and can be consistent with Planck observation for $n=4$~\cite{Planck:2018jri,Dimopoulos:2020kol}.}
Since such a modification does not affect the post-inflationary dynamics, below we simply use the inflaton potential of the form (\ref{V}).

For a fixed value of $n$, the potential contains two dimensionful parameters $\Lambda$ and $v$, but one of which is determined by the condition that the density perturbation $P_\zeta=2.2\times 10^{-9}$ is reproduced. 
We choose $v/M_{\rm Pl}$ as a free parameter and $\Lambda$ is derived as
\begin{align}
	\Lambda \simeq \begin{cases}
    \displaystyle 9\times 10^{14}\,{\rm GeV} \left(\frac{v}{M_{\rm Pl}}\right)\left(\frac{60}{N}\right)^{\frac{3}{4}}  & (n=4)\\
    \displaystyle 1\times 10^{15}\,{\rm GeV}\left(\frac{v}{M_{\rm Pl}}\right)^{\frac{3}{4}}\left(\frac{60}{N}\right)^{\frac{5}{8}}  & (n=6)
    \end{cases}
\end{align}
Other useful quantities are the Hubble parameter around the end of inflation, $H_{\rm e}$, and the inflaton mass $m_\phi$ around the potential minimum. We define $H_{\rm e} \equiv \Lambda^2/(\sqrt3 M_{\rm Pl})$ and it is given by
\begin{align}
	H_{\rm e} \simeq \begin{cases}
    \displaystyle 2\times 10^{11}\,{\rm GeV} \left(\frac{v}{M_{\rm Pl}}\right)^2\left(\frac{60}{N}\right)^{\frac{3}{2}}  & (n=4)\\
    \displaystyle 4\times 10^{11}\,{\rm GeV}\left(\frac{v}{M_{\rm Pl}}\right)^{\frac{3}{2}}\left(\frac{60}{N}\right)^{\frac{5}{4}}  & (n=6)
    \end{cases}.
\end{align}
The inflaton mass $m_\phi$ is defined as $V \simeq \frac{1}{2}m_\phi^2(\delta\phi)^2$ when $\phi$ is expanded around the minimum as $\phi\simeq v+\delta\phi$, and it is evaluated as
\begin{align}
	m_{\phi}= \frac{\sqrt{2}n \Lambda^2}{v} = \frac{\sqrt{6}n M_{\rm Pl}}{v} H_{\rm e} \simeq \begin{cases}
    \displaystyle 2\times 10^{12}\,{\rm GeV} \left(\frac{v}{M_{\rm Pl}}\right)\left(\frac{60}{N}\right)^{\frac{3}{2}}  & (n=4)\\
    \displaystyle 6\times 10^{12}\,{\rm GeV}\left(\frac{v}{M_{\rm Pl}}\right)^{\frac{3}{2}}\left(\frac{60}{N}\right)^{\frac{5}{4}}  & (n=6)
    \end{cases}.
\end{align}
Note that $m_\phi \gg H_{\rm e}$ for $v/M_{\rm Pl}\ll 1$.
One can clearly see that the inflation scale becomes lower for smaller $v$. Below we consider the case of $v=(1,0.1,0.01)\times M_{\rm Pl}$ as representative values. 

Another free parameter that characterizes the inflaton dynamics is the reheating temperature $T_{\rm RH}$. 
It is determined by the inflaton interaction strength with other SM particles, which do not appear in the inflation model definition (\ref{V}), so we regard $T_{\rm RH}$ as a free parameter.
While $T_{\rm RH}$ affects the inflationary predictions only logarithmically through the change of $N$, it significantly affects the GW spectrum.



\section{Gravitational wave spectrum}
\label{sec:gw_def}

Throughout this paper we consider GWs as tensor perturbation of the spatially flat Friedmann-Lemaitre-Robertson-Walker (FLRW) metric
\begin{align}
	ds^2=-dt^2 + a^2(t)(\delta_{ij} + h_{ij}(t,\vec x)) dx^idx^j.
\end{align}
It is expressed in the momentum space as
\begin{align}
	h_{ij}(t,\vec x) = \sum_{\lambda=+,\times} \int\frac{d^3k}{(2\pi)^3} \left[ a_{k,\lambda} h_{k,\lambda}(t) + a^\dagger_{-k,\lambda} h^*_{k,\lambda}(t)  \right]e^{i\vec k\cdot\vec x} e^\lambda_{ij},
    \label{eq:gravmomentumSpace}
\end{align}
where we introduced creation and annihilation operators $a_{k,\lambda}^{(\dagger)}$. We quantize gravitational field by imposing the canonical commutation relation $[ a_{k,\lambda}, a^\dagger_{k',\lambda'}  ] = (2\pi)^3 \delta(\vec k-\vec k') \delta_{\lambda\lambda'}$. There are two graviton polarizations satisfying the relation $e_{ij}^\lambda e_{ij}^{*\lambda'} =  2\delta_{\lambda\lambda'}$. 
The tensor perturbation follows the equation of motion
\begin{align}
	\ddot h_{k,\lambda} + 3 H \dot h_{k,\lambda} + \left(\frac{k}{a}\right)^2 h_{k,\lambda} =0.
    \label{eq:gravitonEOM}
\end{align}
Eq.~\eqref{eq:gravitonEOM} can be transformed into a time dependent frequency harmonic oscillator using the change of variables $\widetilde h_{k,\lambda} \equiv a M_{\rm pl}h_{k,\lambda}/\sqrt{2} $ and $\tau = \int \frac{dt}{a}$, i.e. by going to the conformal time. The oscillator equation is
\begin{align}
	\widetilde h_{k,\lambda}'' + \omega_k^2 \widetilde h_{k,\lambda} = 0,~~~~~~
	 \omega_k^2 = k^2 - \frac{a''}{a},
     \label{eq:gravitonOscillator}
\end{align}
where $\prime=d/{d\tau}$ is the derivative with respect to $\tau$. 


A classical problem of general relativity is the absence of a universal notion of the energy of the gravitational field. There are many nonequivalent definitions of the energy of a GW which agree in the deep sub-horizon regime, but differ in the super-horizon regime. As we are dealing with both of the regimes in our calculations, we are going to use a definition using the GW Hamiltonian which naturally covers both cases. The conformal time Einstein-Hilbert action expanded to second order in $h_{ij}$ becomes
\begin{equation}
    S_{\rm EH}^{(2)}=\frac{M_{\rm Pl}^2}{8}\int d\tau d^3xa^2\left(h_{ij}^\prime h_{ij}^\prime-\partial_l h_{ij} \partial_l h_{ij}\right).
\end{equation}
The conjugate momentum is
\begin{equation}
    \Pi_{ij}=\frac{\partial\mathcal{L}}{\partial h_{ij}^\prime}=\frac{M_{\rm Pl}^2}{4}a^2 h_{ij}^\prime.
\end{equation}
The Hamiltonian density is thus
\begin{equation}
    \mathcal{H}= h_{ij}^\prime\Pi_{ij}-\mathcal{L}=\frac{M_{\rm Pl}^2}{8}a^2\left( h_{ij}^\prime h_{ij}^\prime+\partial_l h_{ij} \partial_l h_{ij}\right).
    \label{eq:gwHamiltonian}
\end{equation}
In the momentum space \eqref{eq:gwHamiltonian} gives us the Hamiltonian
\begin{equation}
    H(\tau)=\frac{M_{\rm Pl}^2}{4}\sum_{\lambda=+,\times}\int\frac{d^3 k}{(2\pi)^3}a^2\left(\vert h_{\lambda,k}^\prime\vert^2+k^2\vert h_{\lambda,k}\vert^2\right).
\end{equation}
The classical GW energy density inside a comoving volume $V$ can then be written as
\begin{equation}
    \rho_{h}=\frac{H(\tau)}{a^4V}
\end{equation}
and in the infinite volume case we obtain
\begin{equation}
    \rho_{h}=\frac{M_{\rm Pl}^2}{4a^2}\sum_{\lambda=+,\times}\int\frac{d^3 k}{(2\pi)^3}\left(\vert h_{\lambda,k}^\prime\vert^2+k^2\vert h_{\lambda,k}\vert^2\right).
\end{equation}
The logarithmic energy density spectrum becomes
\begin{equation}
    \frac{d\rho_h}{d \ln k}=\frac{k^3M_{\rm Pl}^2}{8\pi^2 a^2}\sum_{\lambda=+,\times}\left(\vert h_{\lambda,k}^\prime\vert^2+k^2\vert h_{\lambda,k}\vert^2\right).
\end{equation}

To obtain the vacuum we solve \eqref{eq:gravitonEOM} in the far past $\tau\rightarrow-\infty$. All modes eventually become deeply subhorizon and the solution can be approximated in the WKB formalism as
\begin{equation}
    \widetilde h_{k,\lambda}(\tau)=\frac{1}{\sqrt{2\omega_k(\tau)}}e^{-i\int_{\tilde{\tau}}\omega_k(\tilde{\tau})d\tilde{\tau}},
\end{equation}
which in the deep subhorizon regime $k^2\gg\frac{a^{\prime\prime}}{a}$ becomes
\begin{align}
	\widetilde h_{k,\lambda}(\tau) = \frac{1}{\sqrt{2k}} e^{-ik\tau}.
\end{align}
This is called the Bunch-Davies vacuum. 
The renormalized physical GW density in term of the canonical variable $\widetilde h_{\lambda,k}$ is given by
\begin{equation}
    \rho_{h}^{\rm (phys)}=\frac{1}{2a^4}\sum_{\lambda=+,\times}\int\frac{d^3 k}{(2\pi)^3}\left(\vert\widetilde h_{\lambda,k}^\prime-\mathcal{H}\widetilde h_{\lambda,k}\vert^2+k^2\vert\widetilde h_{\lambda,k}\vert^2-k\right),~~~\mathcal{H}=\frac{a^\prime}{a}.
\end{equation}
In the deep subhorizon regime the $-\mathcal{H}\widetilde h_{\lambda,k}$ becomes negligible and we can omit it. Whenever we mention GW energy density, we mean this renormalized energy density $\rho_{h}^{\rm (phys)}$ and omit the superscript (phys) for notational simplicity.


GWs are produced in the early universe in several ways as we will see in the following sections.
In order to calculate the GW spectrum in the present universe, we need some conversion formula.
Present day GW frequency is obtained by 
\begin{equation}
    f_0=\frac{k}{2\pi a_0}= f_{\rm R}\frac{T_0}{T_{\rm R}}\left(\frac{g_{*s0}}{g_{*s{\rm R}}}\right)^{1/3},
    \label{f0_R}
\end{equation}
where the index 0 denotes present time values while R denotes the values at the end of reheating. 
This connects the GW frequency at the end of reheating $f_{\rm R}$ with the present frequency. 
The GW energy density spectrum can be written as
\begin{equation}
    \rho_h(k)\equiv\left.\frac{d\rho_h}{d \ln k}\right|_0=\left(\frac{a_{\rm R}}{a_0}\right)^4\left.\frac{d\rho_h}{d \ln k}\right|_{\rm R}.
\end{equation}
Let us define the GW density parameter by
\begin{equation}
    \Omega_{\rm GW}(k)=\frac{\rho_h(k)}{\rho_{\rm crit}},
\end{equation}
where $\rho_{\rm crit}$ is the critical energy density of the universe.
Using comoving entropy conservation and assuming radiation domination at the end of reheating, we can write
\begin{equation}
    \Omega_{\rm GW, 0}(k)=\Omega_{r,0}\frac{g_{*\rm R}}{g_{*0}}\left(\frac{g_{*s0}}{g_{*s \rm R}}\right)^{4/3}\Omega_{\rm GW,R}(k),
    \label{OGW_0_R}
\end{equation}
where $\Omega_{r,0}$ denotes the radiation energy density in the present universe.
Eqs.~(\ref{f0_R}) and (\ref{OGW_0_R}) convert the quantities evaluated at the end of reheating to the present ones.

\section{Gravitational waves from preheating}
\label{sec:gw_p}

\subsection{Preheating}

GWs can be emitted through the nonlinear dynamics of self-interacting scalar fields after inflation, which is known as preheating.
The evolution of the inflaton condensation is governed by the following classical field equation and the Friedmann equation,
\begin{align} \label{eq:inflatonEoM}
    &\ddot{\phi} + 3H \dot\phi - \frac{\nabla^2 \phi}{a^2} + \frac{\partial V}{\partial \phi} = 0,\\
    &3M_{\rm Pl}^2H^2 = \frac{1}{2} \dot{\phi}^2 + \frac{1}{2a^2}(\nabla \phi)^2 + V(\phi).
\end{align}
Right after the end of the inflation, the scalar field can be split into the background value and the fluctuation, $\phi(t,\vec{x}) = \bar\phi(t) + \delta\phi(t,\vec{x})$ and the background field oscillates around the minimum of the inflaton potential. Then, due to the self-coupling, the fluctuation can be amplified through the tachyonic or parametric resonance instability~\cite{Kofman:1994rk,Kofman:1997yn,Felder:2001kt,Felder:2000hj,Desroche:2005yt}. Since the fluctuation can be exponentially enhanced, it can exceed the background value. After that, the system enters fully nonlinear regime and the inflaton field is no longer separated into the coherent component and fluctuations.

For the new inflation model (\ref{V}), there are tachyonic enhancement of inflaton fluctuations since the potential has a negative curvature around the top of the potential.
The mean inflaton spatial fluctuation $\left<\delta\phi^2\right>$ grows as~\cite{Brax:2010ai,Ema:2017rkk}
\begin{align}
    \left<\delta\phi^2\right>\sim m_\phi^2\left(\frac{M_{\rm Pl}}{v}\right)^{\frac{n-2}{n}\left[2j(x_n-1)-1\right]},~~~~~~x_n\equiv \frac{\sqrt{2n(n-1)}}{n-2},
\end{align}
after $j$-th inflaton oscillation.
Thus it is seen that the inflaton dynamics enters in the nonlinear regime after a few oscillations for $v \lesssim 0.1\,M_{\rm Pl}$, while the growth is highly inefficient for $v \sim M_{\rm Pl}$.

GWs can be efficiently emitted in the preheating phase because of the highly inhomogeneous configuration of the scalar field.
One can follow the emission of GWs with the following evolution equation,
\begin{align} \label{eq:GWEoM}
    \ddot{h}_{ij} + 3H\dot{h}_{ij} - \frac{\nabla^2 h_{ij}}{a^2} = \frac{2}{M_{\rm Pl}^2}\Pi_{ij}^{\rm TT}
\end{align}
where the right hand side denotes the transverse-traceless component of the anisotropic stress of the scalar field given by $\Pi_{ij} = \partial_i \phi \partial_j \phi$.

\subsection{Lattice results}

As written above, the inflaton dynamics may become nonlinear soon after the inflation ends for $v \lesssim 0.1M_{\rm Pl}$.
Thus, the perturbative approach is no longer valid and we have to employ the lattice simulation to solve the equations of motion (\ref{eq:inflatonEoM}), (\ref{eq:GWEoM}) and calculate the resulting GW spectrum.
There have been several studies for lattice simulation of new inflation~\cite{Desroche:2005yt,Antusch:2015nla,Ema:2017rkk}, where it has been found that the preheating is really effective for $v\ll M_{\rm Pl}$.
Still direct simulations of the GW production, and the evolution of the equation of state of the universe during the preheating, which is essential for evaluating the present GW spectrum, are lacking.
Also, the transition from the highly nonlinear case ($v\ll M_{\rm Pl}$) to the linear case ($v\sim M_{\rm Pl}$) is unexplored.\footnote{
    Ref.~\cite{Das:2025tzs} only considered the case of $v \gtrsim M_{\rm Pl}$ and found negligibly small preheating GWs.
}
Below we reveal these things.

We performed lattice simulations to solve the equation of motion and the Friedmann equation self-consistently. The number of grids and the initial box size are $N^3 = 2048^3$ and $L = 256/v$ respectively. We take the initial values for the homogeneous mode $\phi_i = 0.1\,v$, $0.3\,v$ for $n=4,6$ respectively at the initial time $t_i = m_\phi^{-1}$. The scale factor is normalized by $a(t_i) = 1$. In addition, we set the initial fluctuations given by the vacuum quantum fluctuations during inflation (see, e.g. \cite{Figueroa:2020rrl}). We approximate the spatial derivative as the second order finite difference and update the fields using the second order leap-frog method

\begin{figure}[tbp]
\centering
\includegraphics[width=0.48\textwidth]{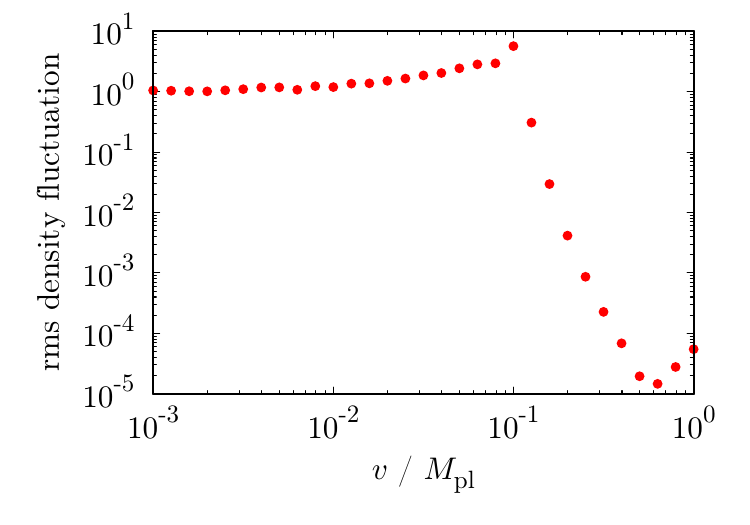}
\includegraphics[width=0.48\textwidth]{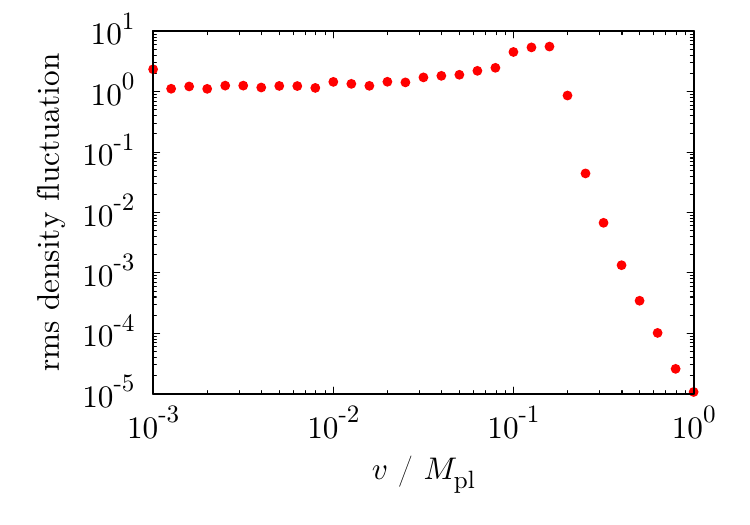}
\caption{\label{fig:growth}The root-mean-square of the density fluctuation at the final time of the simulation for $n=4$ (left) and $6$ (right).}
\end{figure}

\begin{figure}[tbp]
\centering
\includegraphics[width=0.48\textwidth]{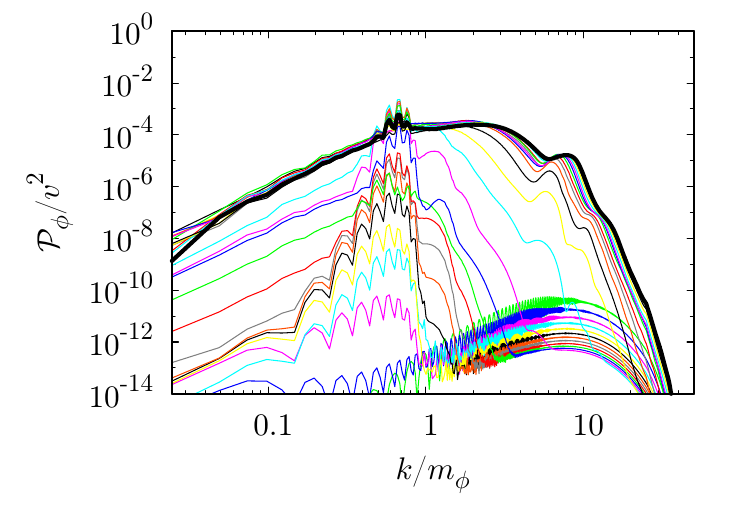}
\includegraphics[width=0.48\textwidth]{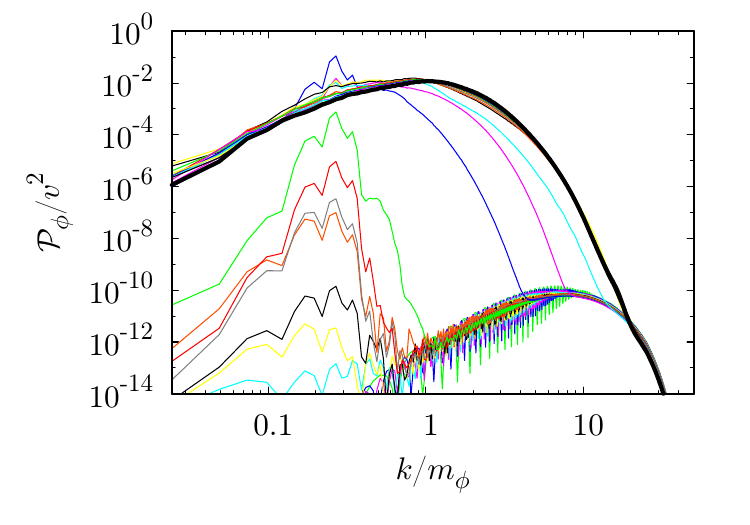}
\includegraphics[width=0.48\textwidth]{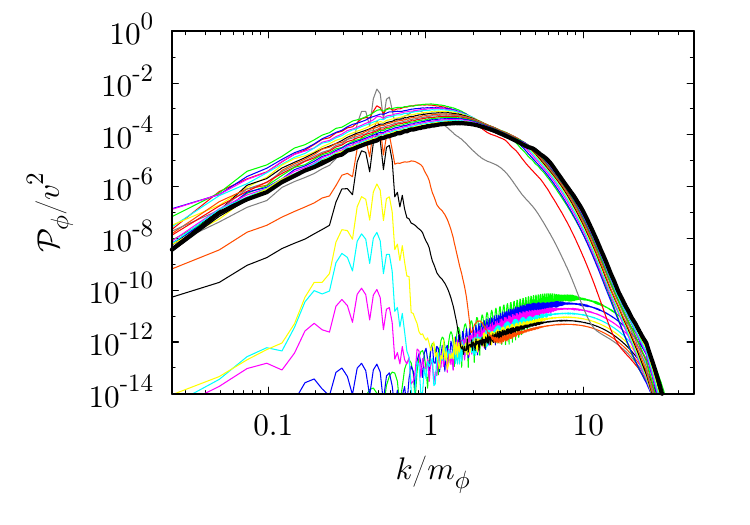}
\includegraphics[width=0.48\textwidth]{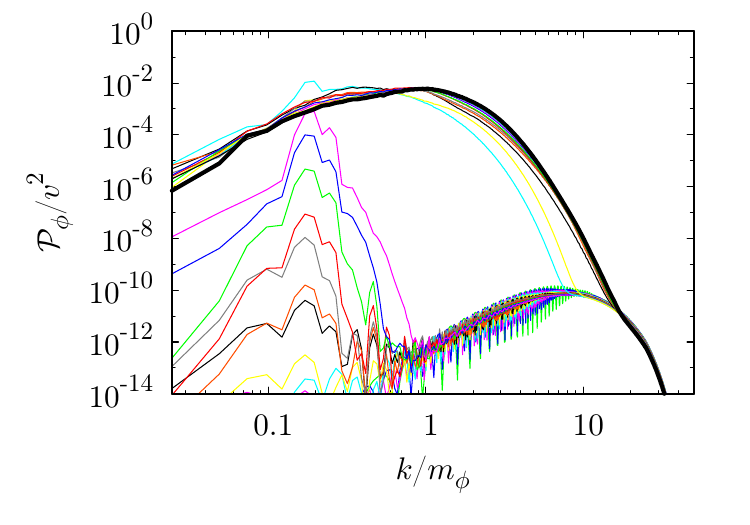}
\caption{\label{fig:spectrum_phi}Time evolution of the spectrum of the scalar field value for $v = 0.1M_{\rm Pl}$ (left), $0.01M_{\rm Pl}$ (right) and $n=4$ (top) and $6$ (bottom).The thick line corresponds to the final time at $mt = 500$ (left), $250$ (right).}
\end{figure}

\begin{figure}[tbp]
\centering
\includegraphics[width=0.48\textwidth]{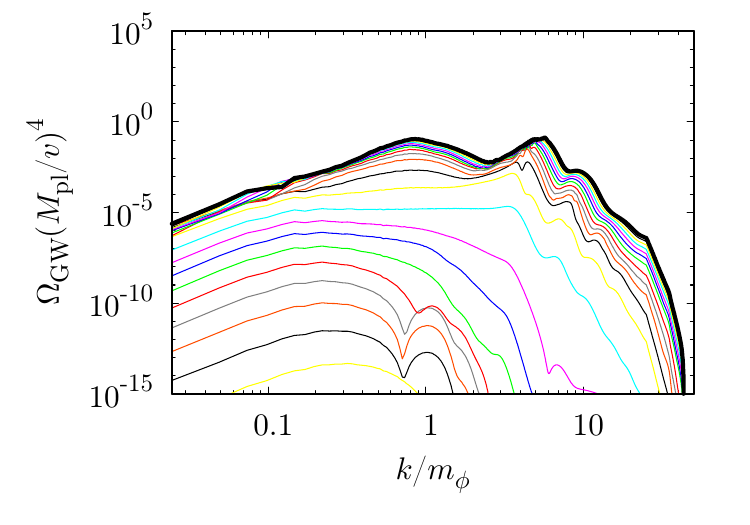}
\includegraphics[width=0.48\textwidth]{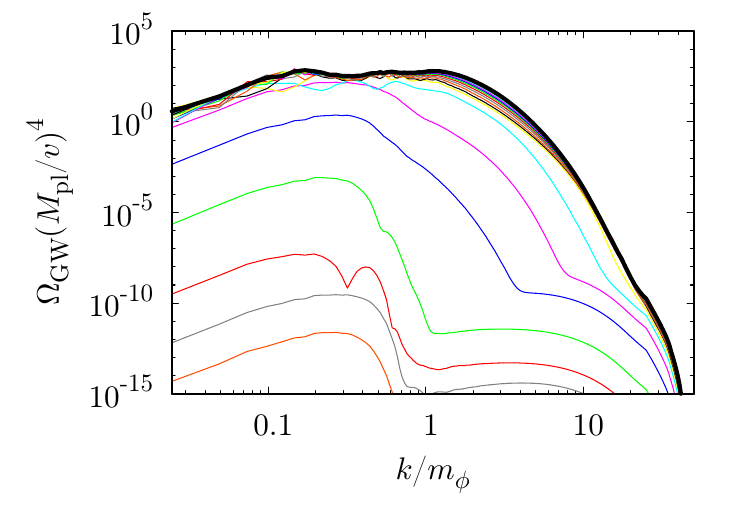}
\includegraphics[width=0.48\textwidth]{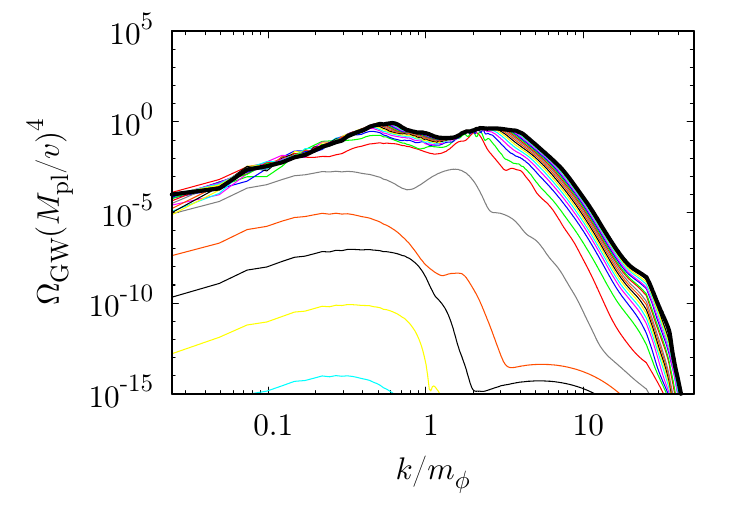}
\includegraphics[width=0.48\textwidth]{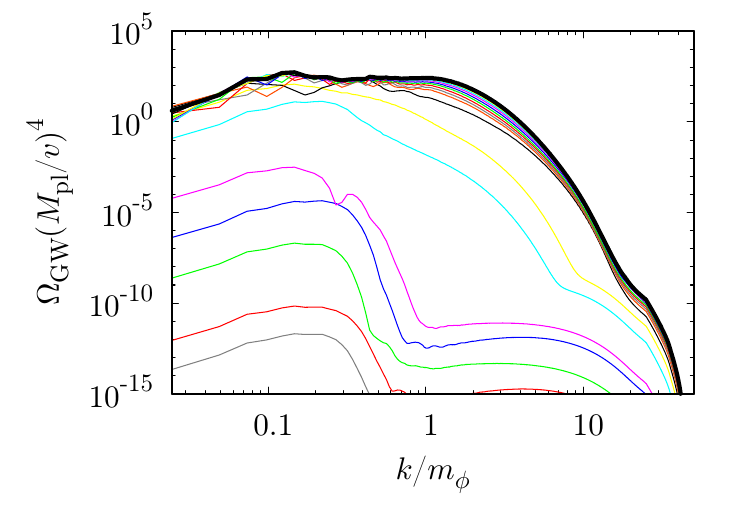}
\caption{\label{fig:spectrum_GW}Time evolution of the spectrum of gravitational waves for $v = 0.1M_{\rm Pl}$ (left), $0.01M_{\rm Pl}$ (right) and $n=4$ (top) and $6$ (bottom). The thick line corresponds to the final time at $mt = 500$ (left), $250$ (right).}
\end{figure}

Fig.~\ref{fig:evolve} shows the evolution of the field value of the inflaton, the root-mean-square of the density contrast, and the equation of state parameter defined by $w = p/\rho$ with $p$ and $\rho$ being respectively the pressure and the energy density of the inflaton.
The numerical results show a significant enhancement of the inflaton fluctuations after several oscillations while the equation of state parameter converges to zero.
Fig.~\ref{fig:growth} shows the density contrast at the final time of the simulation, $mt = 500$, as a function of $v/M_{\rm Pl}$. We have taken $N^3 = 512^3$ and $L = 64/v$ in this calculation. This figure indicates that a significant enhancement in the density fluctuation occurs for $v \lesssim 0.1 M_{\rm Pl}$.
Fig.~\ref{fig:spectrum_phi} and \ref{fig:spectrum_GW} show respectively the evolutions of the spectra of $\phi$ and $\Omega_{\rm GW}$.
They show that a sizable amount of GWs can be emitted with the peak frequency corresponding to the inflaton mass. 
The present-day GW spectra in terms of the current frequency are shown in Figs.~\ref{fig:gw10}-\ref{fig:gw13}.

\begin{figure}[tbp]
\centering
\includegraphics[width=0.48\textwidth]{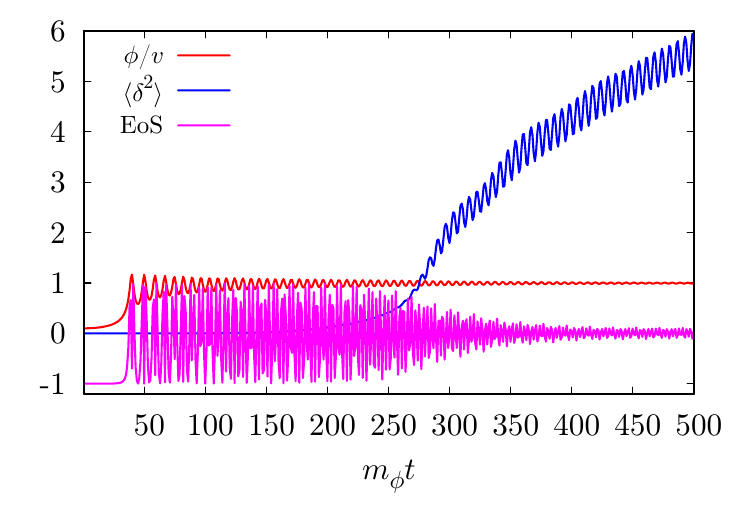}
\includegraphics[width=0.48\textwidth]{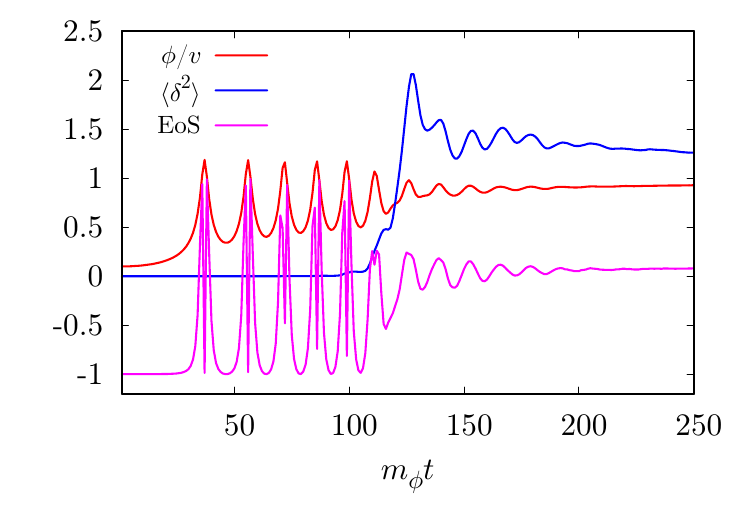}
\includegraphics[width=0.48\textwidth]{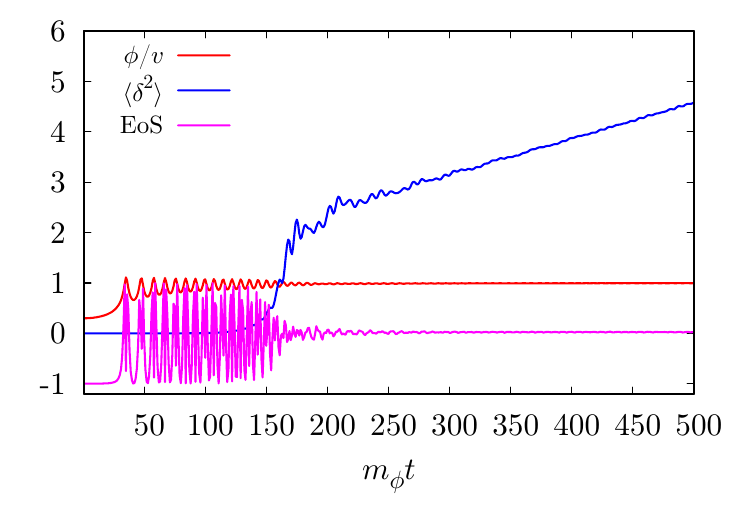}
\includegraphics[width=0.48\textwidth]{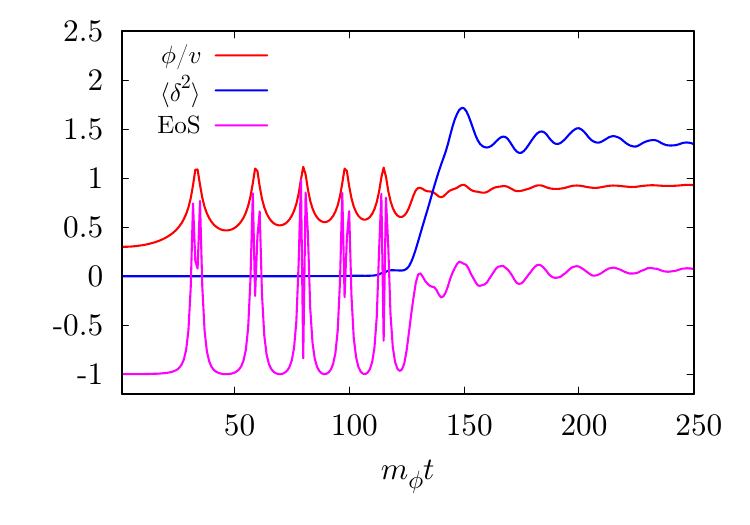}
\caption{\label{fig:evolve}Time evolution of the field value (red), root-mean-square of the density contrast (blue) and the equation of state (magenta) for $v = 0.1M_{\rm Pl}$ (left), $0.01M_{\rm Pl}$ (right) and $n=4$ (top) and $6$ (bottom).}
\label{fig:evolve}
\end{figure}

\section{Gravitational waves from quantum origin}
\label{sec:gw_q}

In this section we study GWs from quantum origin, i.e., those \textit{not} produced by a classical source term in the graviton equation of motion.
Throughout this section, we mostly assume that the inflaton is at rest.
If the preheating happens efficiently, the inflaton fluctuations grow and it gains momentum.
However, as shown in Fig.~\ref{fig:evolve}, still the equation of state of the inflaton remains close to zero on average, meaning that the inflaton is mostly non-relativistic.
Moreover, for relatively high $v$, the preheating is inefficient so that the inflaton is considered to be mostly spatially homogeneous.
In any case, the following calculations remain valid at least as orders-of-magnitude estimation.

\subsection{Primordial gravitational waves}
\label{subsec:hftail}

By ``primordial GWs'' we mean GWs produced just by solving the graviton equation of motion (\ref{eq:gravitonOscillator}) without any source term, under the homogeneous inflaton background with the Bunch-Davies initial condition.
For very low frequency GWs, $\omega_k^2$ in (\ref{eq:gravitonOscillator}) becomes negative during inflation and superhorizon modes are produced~\cite{Starobinsky:1979ty}.
During the reheating era after inflation, $a''/a$ in (\ref{eq:gravitonOscillator}) is a rapidly oscillating and very high frequency GWs are produced~\cite{Ema:2015dka,Ema:2016hlw,Ema:2020ggo} through standard particle production processes~\cite{Dolgov:1989us,Traschen:1990sw,Shtanov:1994ce,Kofman:1997yn}.

\paragraph{Low frequency region: inflationary production}

A well known prediction of slow-roll inflation is an almost flat spectrum of primordial GWs formed by the modes that exit the horizon during inflation and reenter after the reheating. From the slow-roll calculation it follows that the spectrum is given by
\begin{equation}
    \rho_{h}(k)=\frac{k^2}{a^2}\frac{H_e^2}{4\pi^2},
\end{equation}
where $H_e$ denotes the Hubble parameter at the horizon exit, which we assume to be constant during inflation for simplicity.
After the modes reenter the horizon, the energy scales as $a^{-4}$, so we can write
\begin{equation}
    \rho_{h}(k)=\frac{k^2}{a_k^2}\frac{H_e^2}{4\pi^2}\left(\frac{a_k}{a}\right)^4,
\end{equation}
where $a_k$ is the scale factor at the moment of horizon reentry. Denoting $a_e$ the scale factor at the end of inflation and assuming reheating with an equation of state parameter $w$, it holds that
\begin{equation}
    \frac{a_k}{a_e}=\left(\frac{H_e}{H_k}\right)^{\frac{2}{3(1+w)}},~~~H_k=\frac{
k}{a_k}.
\label{eq:Hscaling}
\end{equation}
Using \eqref{eq:Hscaling} we get
\begin{equation}
    \rho_{h}(k)=\frac{H_e^4}{4\pi^2}\left(\frac{a_e}{a}\right)^4\left(\frac{a_e H_e}{k}\right)^{\frac{2(1-3w)}{1+3w}}.
\end{equation}
Modes that reenter during matter domination $w=0$ thus have spectrum scaling as $k^{-2}$, while modes that reenter during radiation domination $w=1/3$ exhibit a flat spectrum~\cite{Nakayama:2008ip,Nakayama:2008wy}.
In terms of $\Omega_{\rm GW}$, by using Eq.~(\ref{OGW_0_R}) we get~\cite{Mudrunka:2026kgm}
\begin{equation}
    \Omega_{\rm GW,0}=\Omega_{r,0}\frac{g_{*\rm R}}{g_{*0}} \left(\frac{g_{*s0}}{g_{*s\rm R}}\right)^{\frac{4}{3}}
	\frac{H_e^2}{12\pi^2 M_{\rm Pl}^2}
	\left(\frac{a_{\rm R}H_{\rm R}}{k}\right)^{\frac{2(1-3w)}{1+3w}}.
\end{equation}
This expression is applicable for modes that experienced super-horizon regime during inflation, i.e., $k\lesssim a_eH_e$.

\paragraph{High frequency region: inflaton annihilation}

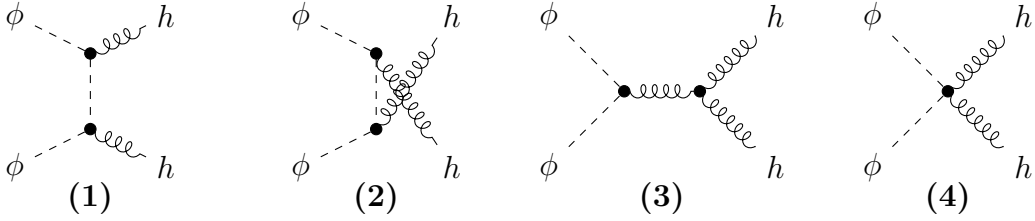
\begin{figure}[htp!]
\centering
\begin{minipage}[c]{0.25\textwidth}
\centering
\begin{tikzpicture}[baseline=(current bounding box.center)]
\begin{feynman}
\vertex (phi1) at (-1.0,1.0) {$\phi$};
\vertex[dot] (v1) at (0,0.5) {};
\vertex[dot] (v2) at (0,-0.5) {};
\vertex (phi2) at (-1.0,-1.0) {$\phi$};
\vertex (g1) at (1.0,1.0) {$h$};
\vertex (g2) at (1.0,-1.0) {$h$};
\diagram*{
  (phi1) -- [scalar] (v1),
  (phi2) -- [scalar] (v2),
  (v1) -- [scalar] (v2),
  (v1) -- [gluon] (g1),
  (v2) -- [gluon] (g2)
};
\end{feynman}
\end{tikzpicture}
\par\vspace{-0.4em}\textbf{(1)}
\end{minipage}\hfill
\begin{minipage}[c]{0.25\textwidth}
\centering
\begin{tikzpicture}[baseline=(current bounding box.center)]
\begin{feynman}
\vertex (phi1) at (-1.0,1.0) {$\phi$};
\vertex[dot] (v1) at (0,0.5) {};
\vertex[dot] (v2) at (0,-0.5) {};
\vertex (phi2) at (-1.0,-1.0) {$\phi$};
\vertex (g1) at (1.0,1.0) {$h$};
\vertex (g2) at (1.0,-1.0) {$h$};
\diagram*{
  (phi1) -- [scalar] (v1),
  (phi2) -- [scalar] (v2),
  (v1) -- [scalar] (v2),
  (v1) -- [gluon] (g2),
  (v2) -- [gluon] (g1)
};
\end{feynman}
\end{tikzpicture}
\par\vspace{-0.4em}\textbf{(2)}
\end{minipage}\hfill
\begin{minipage}[c]{0.25\textwidth}
\centering
\begin{tikzpicture}[baseline=(current bounding box.center)]
\begin{feynman}
\vertex (phi1) at (-1.5,1.0) {$\phi$};
\vertex[dot] (v1) at (-0.5,0) {};
\vertex[dot] (v2) at (0.5,0) {};
\vertex (phi2) at (-1.5,-1.0) {$\phi$};
\vertex (g1) at (1.5,1.0) {$h$};
\vertex (g2) at (1.5,-1.0) {$h$};
\diagram*{
  (phi1) -- [scalar] (v1),
  (phi2) -- [scalar] (v1),
  (v1) -- [gluon] (v2),
  (v2) -- [gluon] (g1),
  (v2) -- [gluon] (g2)
};
\end{feynman}
\end{tikzpicture}
\par\vspace{-0.4em}\textbf{(3)}
\end{minipage}\hfill
\begin{minipage}[c]{0.25\textwidth}
\centering
\begin{tikzpicture}[baseline=(current bounding box.center)]
\begin{feynman}
\vertex (phi1) at (-1.0,1.0) {$\phi$};
\vertex[dot] (v) at (0,0) {};
\vertex (phi2) at (-1.0,-1.0) {$\phi$};
\vertex (g1) at (1.0,1.0) {$h$};
\vertex (g2) at (1.0,-1.0) {$h$};
\diagram*{
  (phi1) -- [scalar] (v),
  (phi2) -- [scalar] (v),
  (v) -- [gluon] (g1),
  (v) -- [gluon] (g2)
};
\end{feynman}
\end{tikzpicture}
\par\vspace{-0.4em}\textbf{(4)}
\end{minipage}
\caption{Diagrams contributing to $\phi\phi\to hh$.}
\label{fig:ann}
\end{figure}

As shown in Refs.~\cite{Ema:2015dka,Ema:2016hlw,Ema:2020ggo}, the inflaton coherent oscillation produces GWs through the process through the oscillating $a''/a$ term in the graviton equation of motion, that can be interpreted as inflaton pair annihilation into gravitons. 
The tree level calculation involves a total of four diagrams shown in Fig.~\ref{fig:ann}.

The inflaton field is by design nonrelativistic, so we can assume both the initial inflaton four momentum to be $(m_\phi,\Vec{0})$. In this case each diagrams yields the following contributions to the amplitude:
\begin{equation}
    \mathcal M^{(1)}_{++}=\mathcal M^{(1)}_{\times\times}=\mathcal M^{(1)}_{\times+}=\mathcal M^{(1)}_{+\times}=0,
\end{equation}
\begin{equation}
    \mathcal M^{(2)}_{++}=\mathcal M^{(2)}_{\times\times}=\mathcal M^{(2)}_{\times+}=\mathcal M^{(2)}_{+\times}=0,
\end{equation}
\begin{equation}
    \mathcal M^{(3)}_{++}=\mathcal M^{(3)}_{\times\times}=-i\frac{3}{4}\kappa^2 m_\phi^2,
    \quad \mathcal M^{(3)}_{\times+}=\mathcal M^{(3)}_{+\times}=0,
\end{equation}
\begin{equation}
    \mathcal M^{(4)}_{++}=\mathcal M^{(4)}_{\times\times}=i\kappa^2 m_\phi^2,\quad \mathcal M^{(4)}_{\times+}=\mathcal M^{(4)}_{+\times}=0,
\end{equation}
where $\kappa\equiv 2/M_{\rm Pl}$ and the subscripts $+$ and $\times$ indicate the graviton polarization.\footnote{
    It is worth mentioning interpretation of these diagrammatic calculations. It is trivial that diagrams (1) and (2) in Fig.~\ref{fig:ann} do not contribute, since the on-shell graviton only couples to spatial gradient of the scalar in the transverse-traceless (TT) gauge and the initial state inflaton momenta are zero. 
    The diagram (4) comes from $\sqrt{-g}\mathcal L_\phi \sim h_{ij}^2 \mathcal L_\phi$, but this is canceled by a part of the contribution from (3), which also contains $\sqrt{-g}\mathcal L_\phi \sim h\mathcal L_\phi$ in the left vertex combined with the term like $\sqrt{-g}R\sim (\Box h) h_{ij}^2$ at the right vertex (note that $\Box h$ is non-vanishing for off-shell graviton). 
    This cancellation ensures that the graviton remains massless. Finally, another part of the diagram (3) with the right vertex from $\sqrt{-g}R\sim h (\partial h_{ij})^2$ remains.
}
Because the inflaton potential also contains self interaction terms, there are further loop diagrams that contribute to the annihilation. However, their contributions are generally negligible compared to the four diagrams above. Summing over the two graviton polarizations and including a factor of $1/4$ to account for the symmetry of the initial and final states we get the production rate to be \cite{Choi:2024ilx}
\begin{equation}
    \Gamma_{\phi\phi\rightarrow hh}=\frac{\rho_{\phi}}{m_{\phi}}\frac{1}{4}\frac{1}{32\pi m_{\phi}^2}\sum_{+\times}\vert \mathcal M\vert ^2=\frac{\rho_{\phi}m_{\phi}}{64\pi M_{\rm Pl}^4}.
\end{equation}
The high frequency graviton production happens at the wave number given by $k=m_\phi a_p$, where $a_p$ denotes the scale factor at the time of production. Taking into account the subsequent redshift, the energy density of the produced gravitons at the end of reheating $a(t)=a_{\rm R}$ is given by
\begin{equation}
    \rho_h (k, a_{\rm R})=\rho_\phi (a_p)\frac{\Gamma_{\phi\phi\rightarrow hh}}{H_p}\left(\frac{a_p}{a_{\rm R}}\right)^4=\frac{9}{64\pi}H_e^3 m_\phi\left(\frac{a_e}{a_{\rm R}}\right)^4\left(\frac{a_e m_\phi}{k}\right)^{1/2},
\end{equation}
where we used the fact that $H\propto a^{-3/2}$ during matter domination and $k = a_p m_\phi$.
Thus the high frequency tail of the spectrum therefore scales as $k^{-1/2}$~\cite{Ema:2015dka,Ema:2016hlw,Ema:2020ggo}. 
In terms of $\Omega_{\rm GW}$, by using Eq.~(\ref{OGW_0_R}) we get
\begin{equation}
    \Omega_{\rm GW,0}(k)=\Omega_{r,0}\frac{g_{*\rm R}}{g_{*0}} \left(\frac{g_{*s0}}{g_{*s\rm R}}\right)^{\frac{4}{3}} \frac{3}{64\pi}\frac{m_\phi H_e^{1/3}H_{\rm R}^{2/3}}{M_{\rm Pl}^2}\left(\frac{a_e m_\phi}{k}\right)^{1/2}.
\end{equation}
This expression is applicable for the modes with $a_em_\phi \lesssim k \lesssim a_{\rm R} m_\phi$.

\paragraph{Intermediate frequency region}

As explained above, the GW spectrum in the low- and high-frequency limits are relatively easy to calculate.
However, it is rather difficult to evaluate the spectrum in the intermediate frequency $a_e H_e\lesssim k \lesssim a_e m_\phi$, at which GWs never experienced super-Hubble radius regime and also the inflaton dynamics cannot be approximated by a periodic oscillation with monomial potential.
In particular, low-scale inflation models such as new inflation often has a hierarchy on the parameter, $H_e\ll m_\phi$, which makes the analytic calculation difficult.
In this intermediate frequency regime we must rely on numerical calculations. This has been first done in detail in Ref.~\cite{Ema:2020ggo} for large field inflation and in Ref.~\cite{Mudrunka:2026kgm} for more general inflation models including small-field ones, which show nontrivial and complicated structure in the GW spectrum.
Later it is also confirmed in Refs.~\cite{Wang:2026ule,Wang:2026pff}.
Detailed calculations are written in Appendix~\ref{app:precise}, where we present new numerical techniques for efficiently calculating GW spectra, which significantly improves the previous work~\cite{Mudrunka:2026kgm}.

\subsection{Bremsstrahlung}

Gravitons emitted by bremsstrahlung during the perturbative inflaton decay process provide another source of high frequency GW spectrum~\cite{Nakayama:2018ptw,Huang:2019lgd,Barman:2023ymn}. 

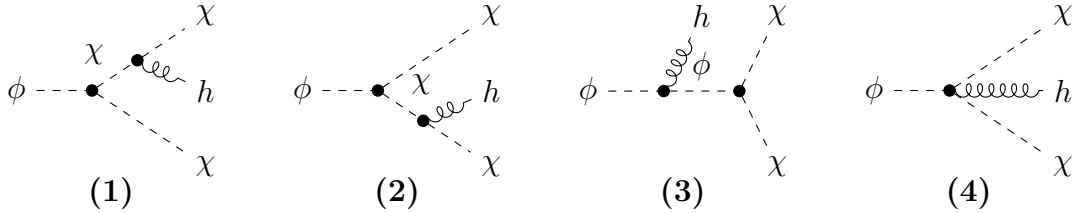
\begin{figure}[htp!]
\centering
\begin{minipage}[c]{0.25\textwidth}
\centering
\begin{tikzpicture}[baseline=(current bounding box.center)]
\begin{feynman}
\vertex (phi) at (-1.0,0) {$\phi$};
\vertex[dot] (v1) at (0,0) {};
\vertex[dot] (v2) at (0.6,0.4) {};
\vertex (varphi1) at (1.5,1.0) {$\chi$};
\vertex (varphi2) at (1.5,-1.0) {$\chi$};
\vertex (h) at (1.5,0) {$h$};
\diagram*{
  (phi) -- [scalar] (v1),
  (v1) -- [scalar, edge label=$\chi$] (v2),
  (v2) -- [scalar] (varphi1),
  (v2) -- [gluon] (h),
  (v1) -- [scalar] (varphi2)
};
\end{feynman}
\end{tikzpicture}
\par\vspace{-0.4em}\textbf{(1)}
\end{minipage}\hfill
\begin{minipage}[c]{0.25\textwidth}
\centering
\begin{tikzpicture}[baseline=(current bounding box.center)]
\begin{feynman}
\vertex (phi) at (-1.0,0) {$\phi$};
\vertex[dot] (v1) at (0,0) {};
\vertex[dot] (v2) at (0.6,-0.4) {};
\vertex (varphi1) at (1.5,1.0) {$\chi$};
\vertex (varphi2) at (1.5,-1.0) {$\chi$};
\vertex (h) at (1.5,0) {$h$};
\diagram*{
  (phi) -- [scalar] (v1),
  (v1) -- [scalar, edge label=$\chi$] (v2),
  (v1) -- [scalar] (varphi1),
  (v2) -- [gluon] (h),
  (v2) -- [scalar] (varphi2)
};
\end{feynman}
\end{tikzpicture}
\par\vspace{-0.4em}\textbf{(2)}
\end{minipage}\hfill
\begin{minipage}[c]{0.25\textwidth}
\centering
\begin{tikzpicture}[baseline=(current bounding box.center)]
\begin{feynman}
\vertex (phi) at (-1.0,0) {$\phi$};
\vertex[dot] (v1) at (0,0) {};
\vertex[dot] (v2) at (1.0,0) {};
\vertex (varphi1) at (1.5,1.0) {$\chi$};
\vertex (varphi2) at (1.5,-1.0) {$\chi$};
\vertex (h) at (0.5,1.0) {$h$};
\diagram*{
  (phi) -- [scalar] (v1),
  (v1) -- [scalar, edge label=$\phi$] (v2),
  (v2) -- [scalar] (varphi1),
  (v1) -- [gluon] (h),
  (v2) -- [scalar] (varphi2)
};
\end{feynman}
\end{tikzpicture}
\par\vspace{-0.4em}\textbf{(3)}
\end{minipage}\hfill
\begin{minipage}[c]{0.25\textwidth}
\centering
\begin{tikzpicture}[baseline=(current bounding box.center)]
\begin{feynman}
\vertex (phi) at (-1.0,0) {$\phi$};
\vertex[dot] (v1) at (0,0) {};
\vertex (varphi1) at (1.5,1.0) {$\chi$};
\vertex (varphi2) at (1.5,-1.0) {$\chi$};
\vertex (h) at (1.5,0) {$h$};
\diagram*{
  (phi) -- [scalar] (v1),
  (v1) -- [scalar] (varphi1),
  (v1) -- [gluon] (h),
  (v1) -- [scalar] (varphi2)
};
\end{feynman}
\end{tikzpicture}
\par\vspace{-0.4em}\textbf{(4)}
\end{minipage}\hfill
\caption{Bremsstrahlung production of the graviton for the process $\phi\to \chi\chi h$.}
\label{fig:brem}
\end{figure}

For concreteness, we consider inflaton-Higgs three point interactions
\begin{equation}
    \mathcal L_{\phi H^\dagger H}=\lambda(\phi-\phi_0) H^\dagger H,~~~H=\frac{1}{\sqrt{2}}\begin{pmatrix}
        \chi_0+i\chi_1 \\ \chi_2+i\chi_3,
    \end{pmatrix}
    \label{3point}
\end{equation}
which contains a total of four real fields with interactions of the type $\frac{\lambda}{2}\phi\chi\chi$. There are in total four diagrams that contribute to it, as shown in Fig.~\ref{fig:brem}.
The resulting contributions to the amplitude per Higgs component are 
\begin{equation}
             \mathcal M^{(1)}=i\frac{\lambda\kappa}{2}\frac{p_\mu p_\nu \epsilon^{\mu\nu}}{p\cdot k},
\end{equation}
\begin{equation}
             \mathcal M^{(2)}=i\frac{\lambda\kappa}{2}\frac{q_\mu q_\nu \epsilon^{\mu\nu}}{m_{\phi}E_k-p\cdot k},
\end{equation}
\begin{equation}
             \mathcal M^{(3)}=0,~~~~\mathcal M^{(4)}=0.
\end{equation}
Again, vanishing of the diagram (3) is understood as a result of the zero momentum of the initial inflaton since the on-shell graviton only couples to spatial gradient of the inflaton in the TT gauge. Diagram (4) also trivially vanishes since the vertex comes from $\sqrt{-g}\mathcal L_{\phi\chi\chi} \sim h \mathcal L_{\phi\chi\chi}$ but the trace $h$ for on-shell graviton vanishes in the TT gauge.
Thus only diagrams (1) and (2) are relevant.
Then the three body decay rate of the inflaton including one graviton final state is given by\footnote{We have summed over the 4 real Higgs components as well as introduced a factor of $1/2$ to account for the symmetry between the two outgoing particles.}
\begin{align}
    \frac{d\Gamma_{1\rightarrow 3}}{d\ln E} = \frac{\lambda^2 m_\phi}{64\pi^3 M_{\rm Pl}^2}\left(1-\frac{2E}{m_\phi}\right)^2.
\end{align}
where $E$ denotes the graviton energy.

For this process the Boltzmann equation for the graviton is written as~\cite{Barman:2023ymn}
\begin{equation}
    \dot{\rho}_h+4H\rho_h=\int_0^{E_{\rm max}}\rho_\phi\frac{E}{m_\phi}\frac{d \Gamma_{1\rightarrow3}}{dE}dE.
\end{equation}
The fraction $E/m_\phi$ corresponds to the part of the total energy injected into the graviton during the decay. $E_{\rm max}=m_\phi/2$ in this case. Taking the derivative with respect to $E$ on both sides and assuming instantaneous reheating, the approximate solution is
\begin{equation}
    \frac{d\rho_h(t_{\rm R})}{d \ln E}\approx\frac{\rho_\phi E}{\Gamma_{\rm tot} m_\phi}\frac{d \Gamma_{1\rightarrow3}}{d \ln E},
\end{equation}
where the right-hand side is evaluated at $t_{\rm R}$ and $\Gamma_{\rm tot}= \lambda^2/(8\pi m_\phi)$. Then, from Eq.~(\ref{OGW_0_R}), the GW density at the present time is given by
\begin{equation}
    \Omega_{\rm GW}^{\rm (brems)}(k)\simeq\Omega_{r,0}\frac{g_{*\rm R}}{g_{*0}}\left(\frac{g_{*s0}}{g_{*s\rm R}}\right)^{4/3} \frac{m_\phi^2}{8\pi^2 M_{\rm Pl}^2} x(1-2x)^2,~~~~~~x\equiv \frac{k}{m_\phi a_{\rm R}}.
\end{equation}
Note that the unknown coupling constant $\lambda$ is canceled out in this expression.\footnote{
    Recently it is shown that the GW spectrum of the low frequency tail changes from $k$ to $k^3$~\cite{Blas:2026yqb,Mukaida:2026bdh}.
}

\subsection{Inflaton-particle scattering}
\label{sec:scatter}

SM particles are gradually produced and form thermal bath during the reheating process.
It has been pointed out that scatterings between the inflaton and SM particles provide another source of high frequency GWs~\cite{Xu:2024fjl,Bernal:2025lxp,Xu:2025wjq}.

\begin{figure}[htp!]
\centering
\begin{minipage}[c]{0.25\textwidth}
\centering
\begin{tikzpicture}[baseline=(current bounding box.center)]
\begin{feynman}
\vertex (phi1) at (-1.5,1.0) {$\phi$};
\vertex[dot] (v1) at (0,0.5) {};
\vertex[dot] (v2) at (0,-0.5) {};
\vertex (phi2) at (-1.5,-1.0) {$\chi$};
\vertex (g1) at (1.5,1.0) {$h$};
\vertex (g2) at (1.5,-1.0) {$\chi$};
\diagram*{
  (phi1) -- [scalar, momentum=\(l\)] (v1),
  (phi2) -- [scalar, momentum=\(p\)] (v2),
  (v1) -- [scalar, edge label = $\phi$] (v2),
  (v1) -- [gluon, momentum=\(k\)] (g1),
  (v2) -- [scalar, momentum=\(q\)] (g2)
};
\end{feynman}
\end{tikzpicture}
\par\vspace{-0.4em}\textbf{(1)}
\end{minipage}\hfill
\begin{minipage}[c]{0.25\textwidth}
\centering
\begin{tikzpicture}[baseline=(current bounding box.center)]
\begin{feynman}
\vertex (phi1) at (-1.0,1.0) {$\chi$};
\vertex[dot] (v1) at (0,0.5) {};
\vertex[dot] (v2) at (0,-0.5) {};
\vertex (phi2) at (-1.0,-1.0) {$\phi$};
\vertex (g1) at (1.0,1.0) {$h$};
\vertex (g2) at (1.0,-1.0) {$\chi$};
\diagram*{
  (phi1) -- [scalar] (v1),
  (phi2) -- [scalar] (v2),
  (v1) -- [scalar, edge label = $\chi$] (v2),
  (v1) -- [gluon] (g1),
  (v2) -- [scalar] (g2)
};
\end{feynman}
\end{tikzpicture}
\par\vspace{-0.4em}\textbf{(2)}
\end{minipage}\hfill
\begin{minipage}[c]{0.25\textwidth}
\centering
\begin{tikzpicture}[baseline=(current bounding box.center)]
\begin{feynman}
\vertex (phi1) at (-1.5,1.0) {$\phi$};
\vertex[dot] (v1) at (-0.5,0) {};
\vertex[dot] (v2) at (0.5,0) {};
\vertex (phi2) at (-1.5,-1.0) {$\chi$};
\vertex (g1) at (1.5,1.0) {$h$};
\vertex (g2) at (1.5,-1.0) {$\chi$};
\diagram*{
  (phi1) -- [scalar] (v1),
  (phi2) -- [scalar] (v1),
  (v1) -- [scalar, edge label=$\chi$] (v2),
  (v2) -- [gluon] (g1),
  (v2) -- [scalar] (g2)
};
\end{feynman}
\end{tikzpicture}
\par\vspace{-0.4em}\textbf{(3)}
\end{minipage}\hfill
\begin{minipage}[c]{0.25\textwidth}
\centering
\begin{tikzpicture}[baseline=(current bounding box.center)]
\begin{feynman}
\vertex (phi1) at (-1.0,1.0) {$\phi$};
\vertex[dot] (v) at (0,0) {};
\vertex (phi2) at (-1.0,-1.0) {$\chi$};
\vertex (g1) at (1.0,1.0) {$h$};
\vertex (g2) at (1.0,-1.0) {$\chi$};
\diagram*{
  (phi1) -- [scalar] (v),
  (phi2) -- [scalar] (v),
  (v) -- [gluon] (g1),
  (v) -- [scalar] (g2)
};
\end{feynman}
\end{tikzpicture}
\par\vspace{-0.4em}\textbf{(4)}
\end{minipage}
\caption{Diagrams contributing to the process $\phi\chi\to h\chi$.}
\label{fig:scat}
\end{figure}
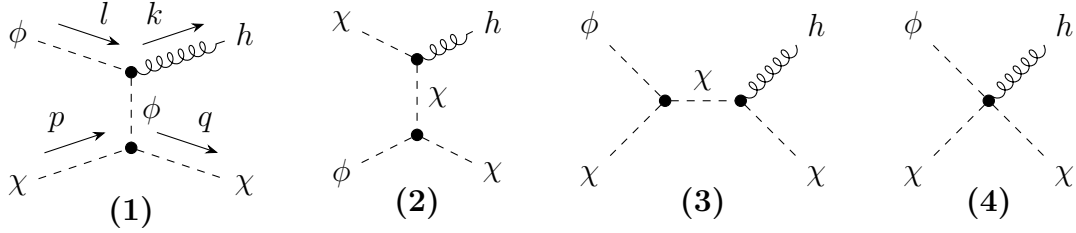

For the case of spin 0 particles coupled through three-point couplings $\mathcal L_{\phi\chi\chi}=\frac{\lambda}{2}\phi\chi\chi$, the process is described by the four diagrams shown in Fig.~\ref{fig:scat}.\footnote{
One could in principle assume the same process with $\chi$ being the inflaton itself. However, in this case the resulting scattering amplitude vanishes under the assumption that the inflaton field is at rest. 
}
The resulting contributions to the amplitude per one Higgs component are (momentum of the incoming inflaton $\phi$ is denoted by $l$, incoming scalar particle $\chi$ by $p$, outgoing scalar field $\chi$ by $q$ and outgoing graviton $h_{\mu\nu}$ by $k$):
\begin{equation}
             \mathcal M^{(1)}=-i\frac{\lambda\kappa}{2}\frac{l_\mu l_\nu \epsilon^{\mu\nu}}{l\cdot k},
\end{equation}
\begin{equation}
             \mathcal M^{(2)}=-i\frac{\lambda\kappa}{2}\frac{p_\mu p_\nu \epsilon^{\mu\nu}}{p\cdot k},
\end{equation}
\begin{equation}
             \mathcal M^{(3)}=i\frac{\lambda\kappa}{2}\frac{q_\mu q_\nu \epsilon^{\mu\nu}}{q\cdot k},
\end{equation}
\begin{equation}
             \mathcal M^{(4)}=0.
\end{equation}
Again the last result $\mathcal M^{(4)}=0$ immediately comes from the fact that the four-point vertex is generated by expanding $\sqrt{-g}\mathcal L_{\phi\chi\chi}$, whose first order perturbation is proportional to the trace $h$ which vanishes for the on-shell graviton in the TT gauge. 
Furthermore, under the assumption that the inflaton is at rest, we also obtain $\mathcal M^{(1)}=0$. This follows from the fact that we can always define the graviton polarization tensor such that $\epsilon_{00}=0$ and therefore
\begin{equation}
    \label{eq:polarizationInflaton}
    l_{\mu}\epsilon^{\mu\nu}=0.
\end{equation}
Thus only diagrams (2) and (3) will contribute to the graviton production. The resulting differential production rate is calculated in detail in Appendix \ref{appendix:scatter}, where we find that the collision term for the Boltzmann equation (\ref{rho_Boltzmann}) is given by
\begin{equation}
    \frac{d\Gamma}{dE}=\frac{\rho_{\phi}}{m_{\phi}^2}\frac{\kappa^{2} \lambda^{2} T^3}{32\pi^3 E}x^3\left[1-\frac{\ln(e^x-1)}{x}\right],~~~~~~x\equiv\frac{2E-m_{\phi}}{2T},
\end{equation}
assuming that $\chi$ particles have thermal distribution with temperature $T$.
As is obvious from Eq.~(\ref{Ek_theta}), the Lorentz kinematics imposes $x\geq 0$. The production rate is peaked around $x\sim 1$.
If $T\ll m_\phi$, this happens at $E\sim m_\phi/2$, while if $T\gg m_\phi$ it corresponds to $E\sim T$.\footnote{
    Our calculation extends that of Ref.~\cite{Xu:2024fjl}, where it has been assumed that $E_{\chi}=E_{\tilde{\chi}}=3T$ and $E=m_{\phi}$. This is possible only when $T\gg m_\phi$.
}
It is easily found that the contribution to the final GW density is peaked around $T\sim T_{\rm RH}$ and the present day spectrum is evaluated as
\begin{align}
        \Omega_{\rm GW}^{\rm (scat)}(k) &\approx \Omega_{r,0}\frac{g_{*\rm R}}{g_{*0}}\left(\frac{g_{*s0}}{g_{*s\rm R}}\right)^{4/3}\frac{T_{\rm RH}^3}{m_{\phi} M_{\rm pl}^2\pi^2}F(x),~~~F(x)=-x^2\ln(1-e^{-x}),
\end{align}
where we assumed that $T_{\rm RH}$ is determined by the perturbative inflaton decay through the three-point coupling $\lambda$ (\ref{3point}).
Note that this is just a contribution from $T\sim T_{\rm RH}$. Depending on the parameters and frequency range, we must integrate over the contributions from all the redshifts. In the numerical calculation presented later, we have taken into account all the contributions by numerically integrating the Boltzmann equation.

Besides the differential spectrum, it is also interesting to see how the total GW energy scales with reheating temperature. We calculate the integrated spectrum 
\begin{equation}
    I(T)=\int_{\frac{m_\phi}{2}}^\infty F(x(E, T))d\ln E=-\int _{0}^\infty \frac{x^2\ln(1-e^{-x})}{x+\mu}dx,~~~~~~\mu\equiv\frac{m_{\phi}}{2T}.
\end{equation}
This factor is calculated as
\begin{equation}
    I(T)\simeq\begin{cases}
        \displaystyle \zeta(3)  & T \gg m_\phi\\
        \displaystyle \frac{2\zeta(4)}{\mu}  &  T \ll m_\phi
    \end{cases},
\end{equation}
where $\zeta$ denotes the Riemann zeta function. 
Therefore, if $T_{\rm RH}\gg m_{\phi}$, we get
\begin{equation}
    \Omega_{\rm GW}^{\rm (scat)}\simeq \Omega_{r,0}\frac{g_{*\rm R}}{g_{*0}}\left(\frac{g_{*s0}}{g_{*s\rm R}}\right)^{4/3} \frac{\zeta(3)}{\pi^2}\frac{T_{\rm RH}^3}{m_{\phi} M_{\rm Pl}^2},
\end{equation}
On the other hand, if $T_{\rm RH} \ll m_{\phi}$, we obtain
\begin{equation}
    \Omega_{\rm GW}^{\rm (scat)}\simeq \Omega_{r,0}\frac{g_{*\rm R}}{g_{*0}}\left(\frac{g_{*s0}}{g_{*s\rm R}}\right)^{4/3}\frac{4\zeta(4)}{\pi^2}\frac{T_{\rm RH}^4}{m_{\phi}^2 M_{\rm Pl}^2}
\end{equation}
For comparison, for the bremsstrahlung GW, we have 
\begin{equation}
    I_{\rm brems}=\int _{0}^{1/2}x(1-2x)^2d\ln x=\frac{1}{6},~~~~~~x\equiv\frac{E}{m_{\phi}},
\end{equation}
so the total energy scales as 
\begin{equation}
     \Omega_{\rm GW}^{\rm (brems)}\simeq\Omega_{r,0}\frac{g_{*\rm R}}{g_{*0}}\left(\frac{g_{*s0}}{g_{*s\rm R}}\right)^{4/3} \frac{m_\phi^2}{48\pi^2 M_{\rm Pl}^2},
\end{equation}
We see that scattering becomes the dominant process once $T_{\rm RH}>m_{\phi}$.

\section{Full gravitational wave spectra}
\label{sec:full}

Now let us combine all the results so far and provide full GW spectra predicted in new inflation model.
We include following ones explained in Sec.~\ref{sec:gw_p} and \ref{sec:gw_q}:
\begin{itemize}
\item GWs from preheating dynamics after inflation
\item Primordial GWs generated during inflation and after inflation
\item Bremsstrahlung GWs during the perturbative reheating
\item GWs produced by scattering of thermal bath particles with the inflaton
\end{itemize}
They are significantly dependent on inflation models and parameters therein.
In the new inflation model (\ref{V}), we choose $v$ as a free parameter, while $\Lambda$ (as well as the inflation scale and the inflaton mass) is fixed by the condition that the correct density perturbation is reproduced.
In addition, we also plotted the following ones:
\begin{itemize}
\item GWs produced by second order scalar perturbation~\cite{Ananda:2006af,Baumann:2007zm,Assadullahi:2009nf,Kohri:2018awv,Inomata:2019zqy}
\item GWs produced by scattering of particles in thermal bath themselves~\cite{Ghiglieri:2015nfa,Ghiglieri:2020mhm,Ringwald:2020ist,Ghiglieri:2022rfp}
\end{itemize}
The former one weekly depends on the scalar spectral index $n_s$, and also on the reheating temperature $T_{\rm RH}$ in the high frequency part, since subhorizon scalar perturbations grow during the matter-dominated era. 
But otherwise it is independent of the inflation model.
The latter one depends only on the reheating temperature.

\begin{figure}
\begin{center}
   \includegraphics[width=0.48\textwidth]{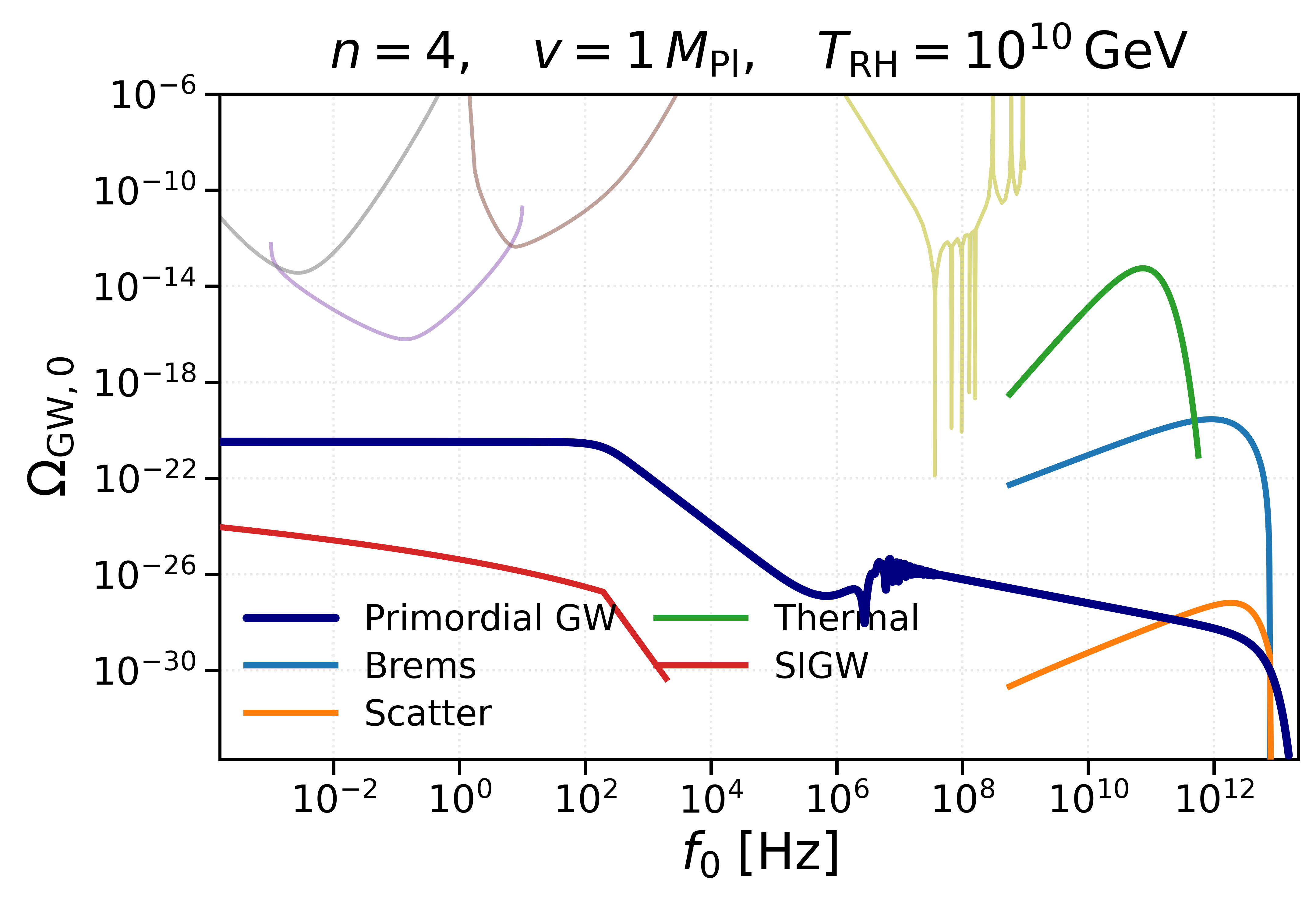}
   \includegraphics[width=0.48\textwidth]{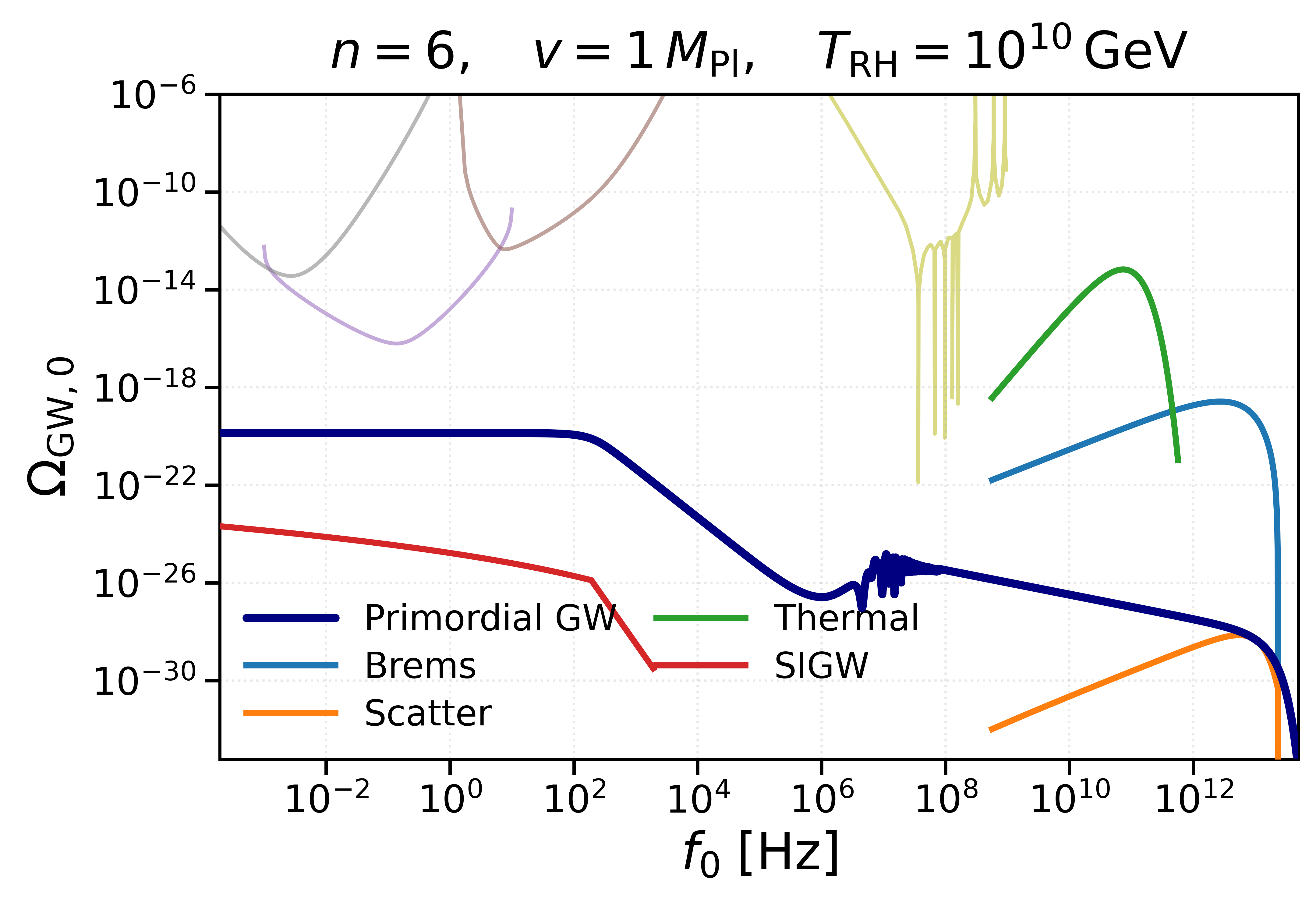}
   \includegraphics[width=0.48\textwidth]{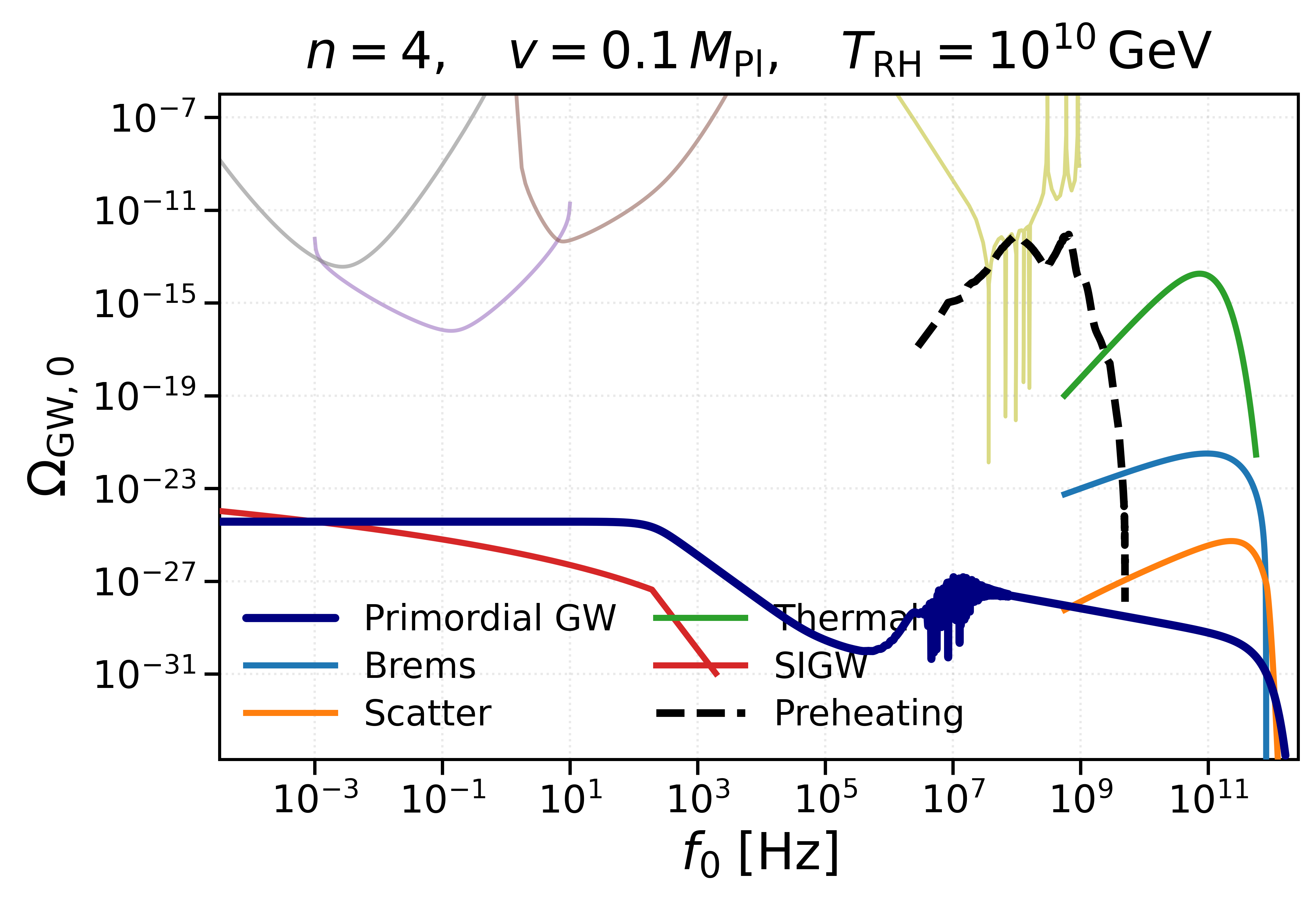}
   \includegraphics[width=0.48\textwidth]{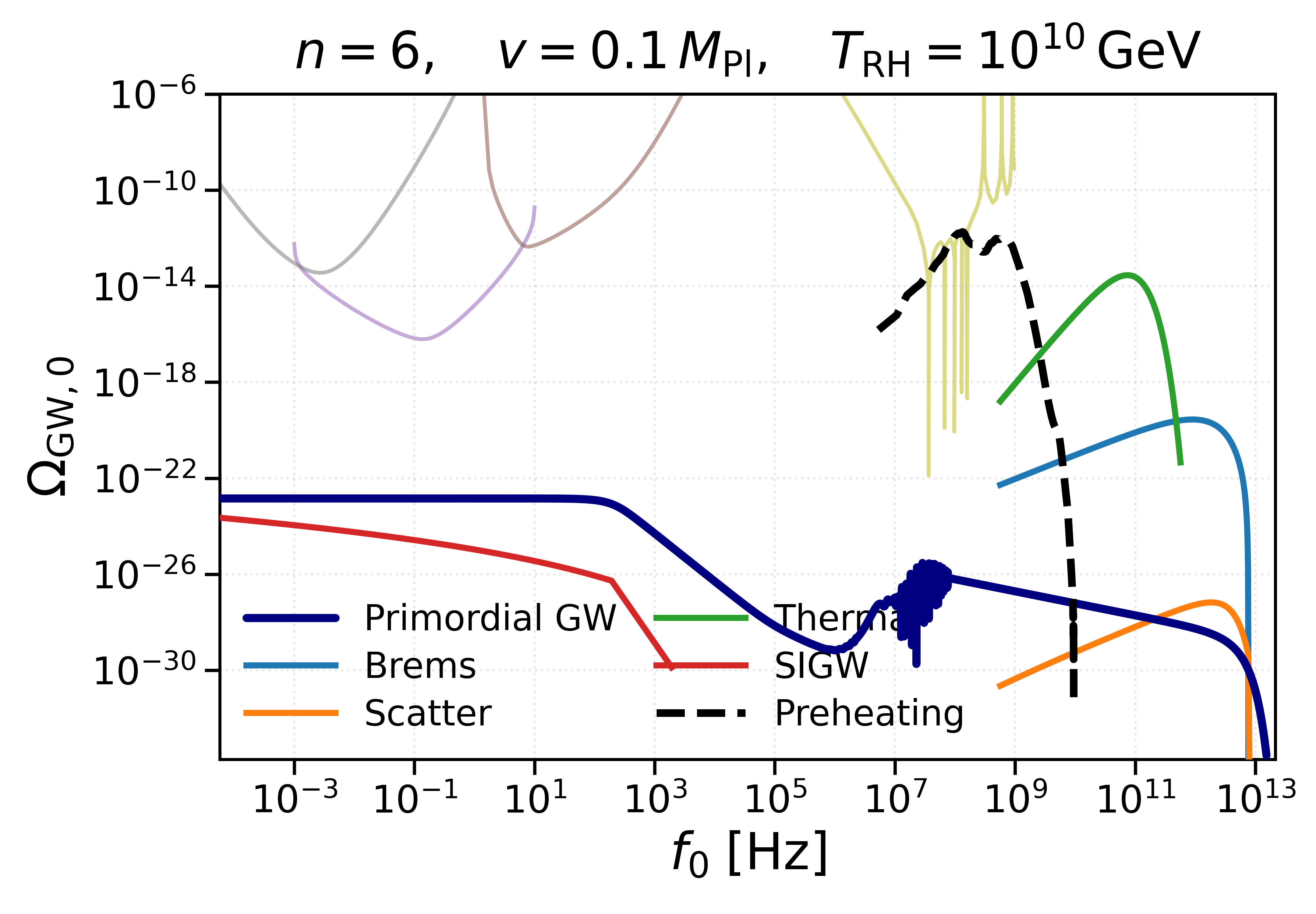}
   \includegraphics[width=0.48\textwidth]{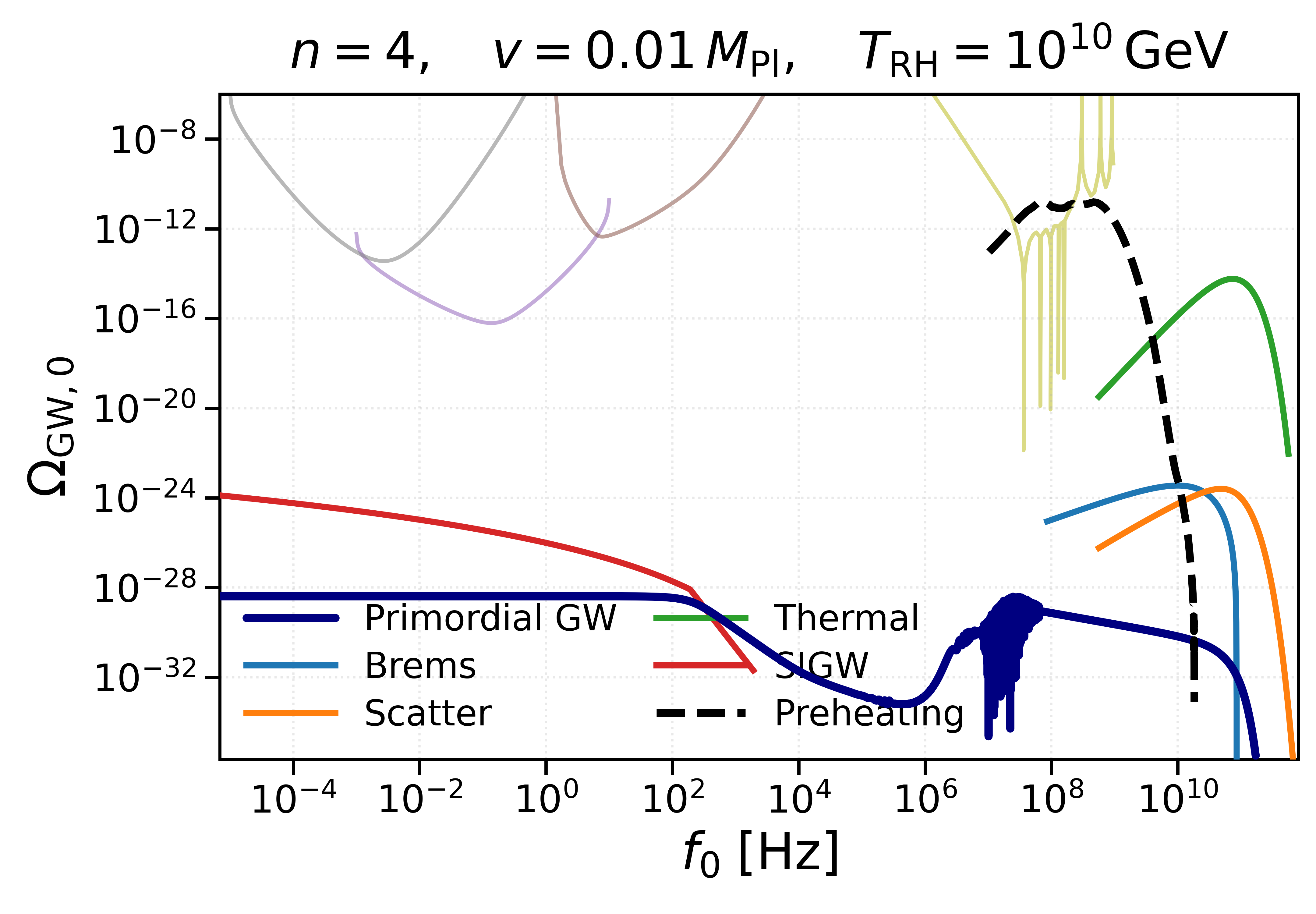}
   \includegraphics[width=0.48\textwidth]{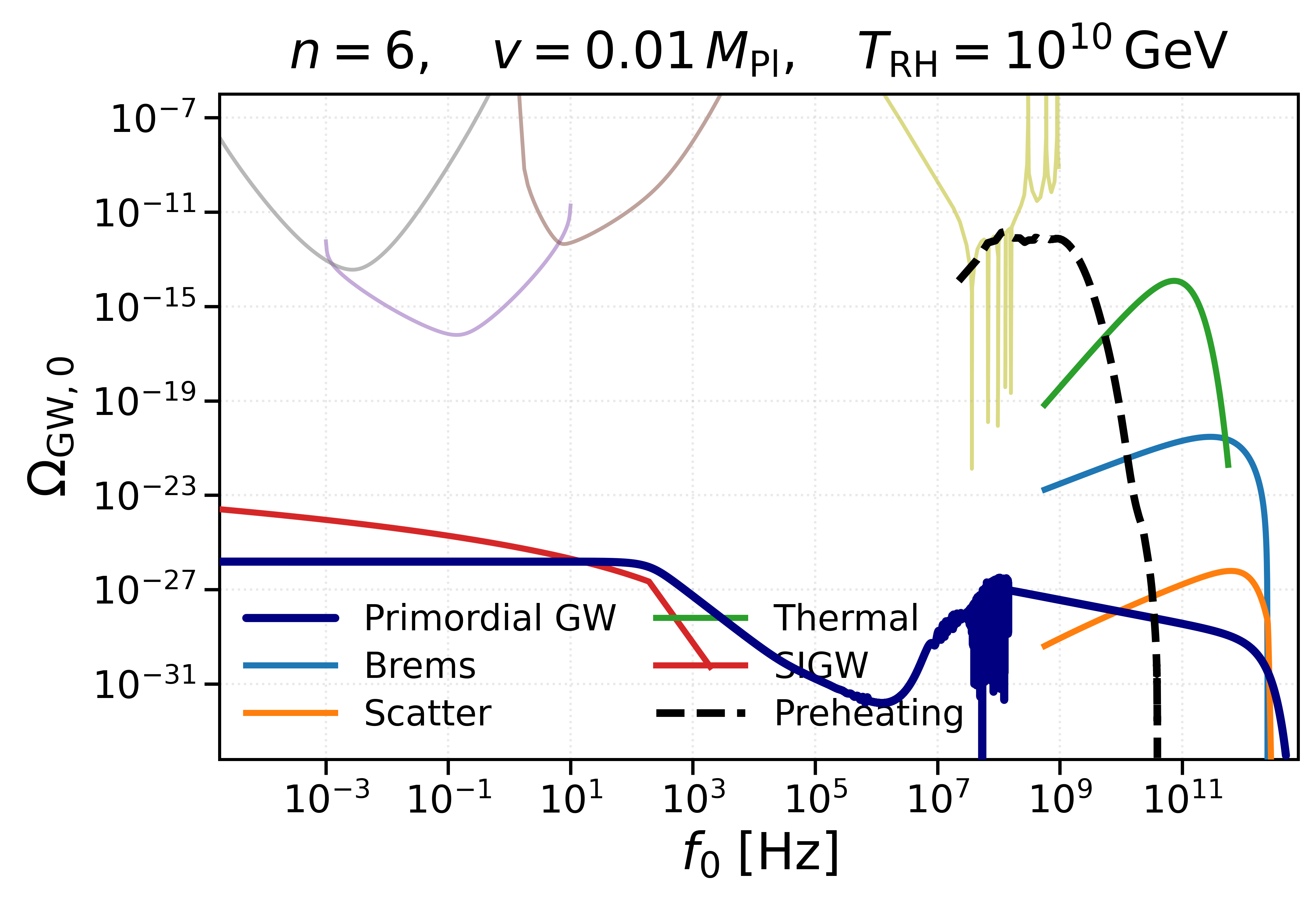}
  \end{center}
  \caption{
Prediction of GW spectra for new inflation model with $n=4$ (left 3 plots) and $n=6$ (right 3 plots) for $T_{\rm RH}=10^{10}\,\rm GeV$. Displayed are the results for $v=M_{\rm Pl}$, $v=0.1M_{\rm Pl}$ and $v=0.01M_{\rm Pl}$, from top to bottom.
``Preheating'' corresponds to GWs produced by the preheating dynamics after inflation.
``Primordial GWs'' represents GWs generated during inflation and reheating, purely induced by the spatially homogeneous inflaton dynamics.
``Brems'' represents bremsstrahlung GWs emitted in the perturbative inflaton decay.
``Scatter'' represents GWs produced by the scattering between the inflaton and thermal particles and ``Thermal'' by the scattering of thermal particles themselves.
``SIGW'' represents scalar-induced GWs generated by the second order curvature perturbation.
The sensitivity curves of future GW observations displayed are SKA, LISA, DECIGO, ET and resonant cavity from left to right.
}
  \label{fig:gw10}
\end{figure}

\begin{figure}
\begin{center}
   \includegraphics[width=0.48\textwidth]{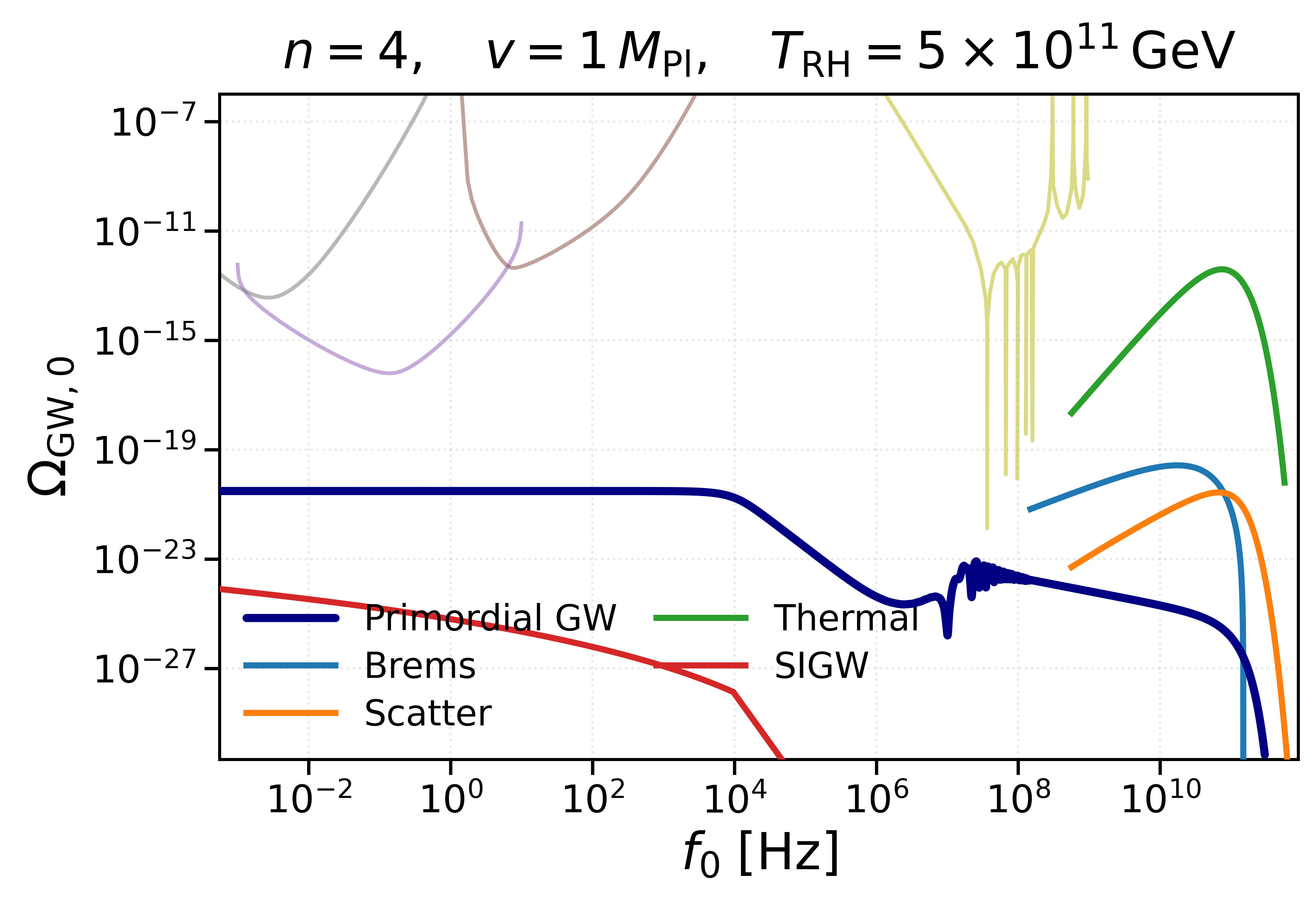}
   \includegraphics[width=0.48\textwidth]{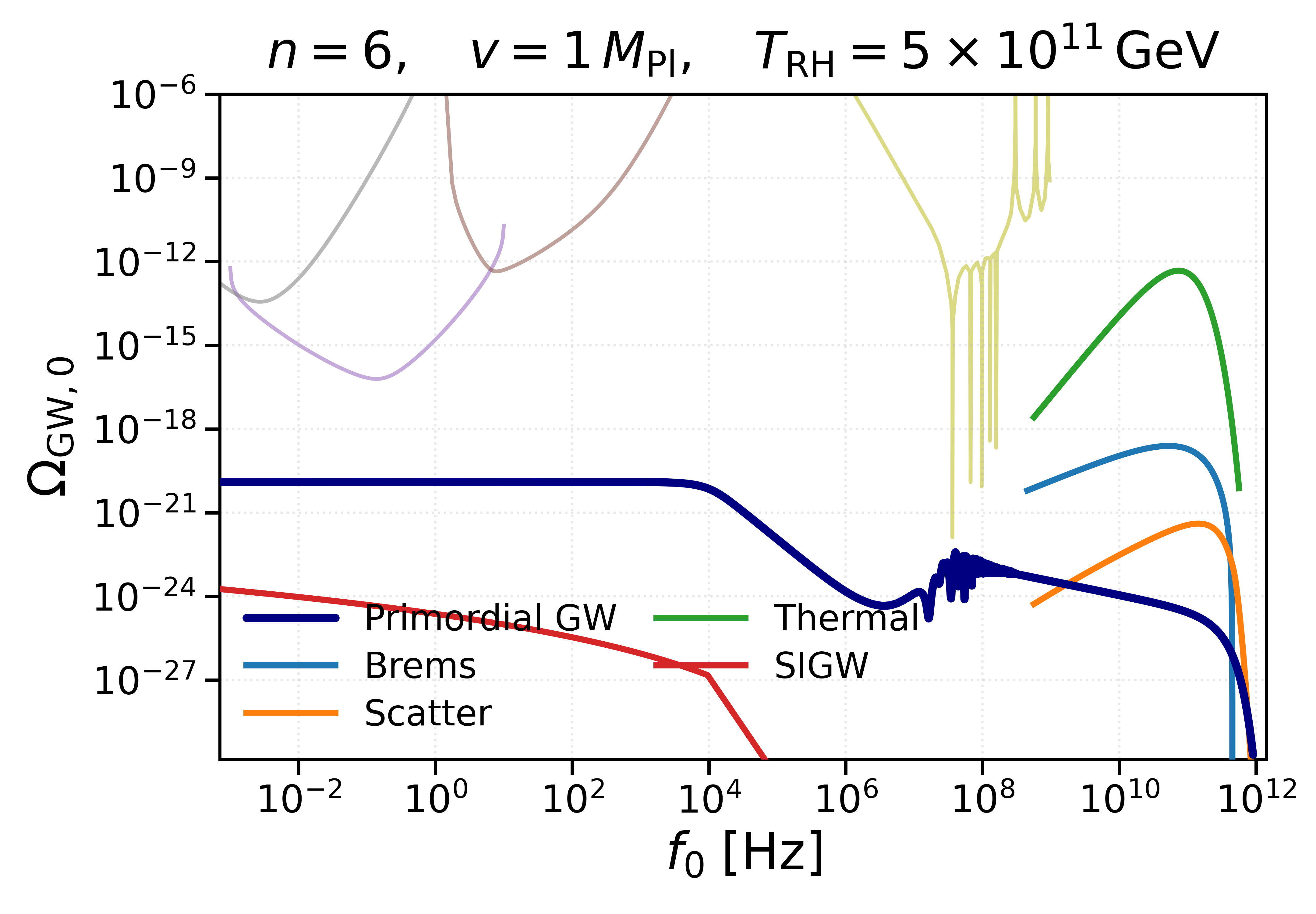}
   \includegraphics[width=0.48\textwidth]{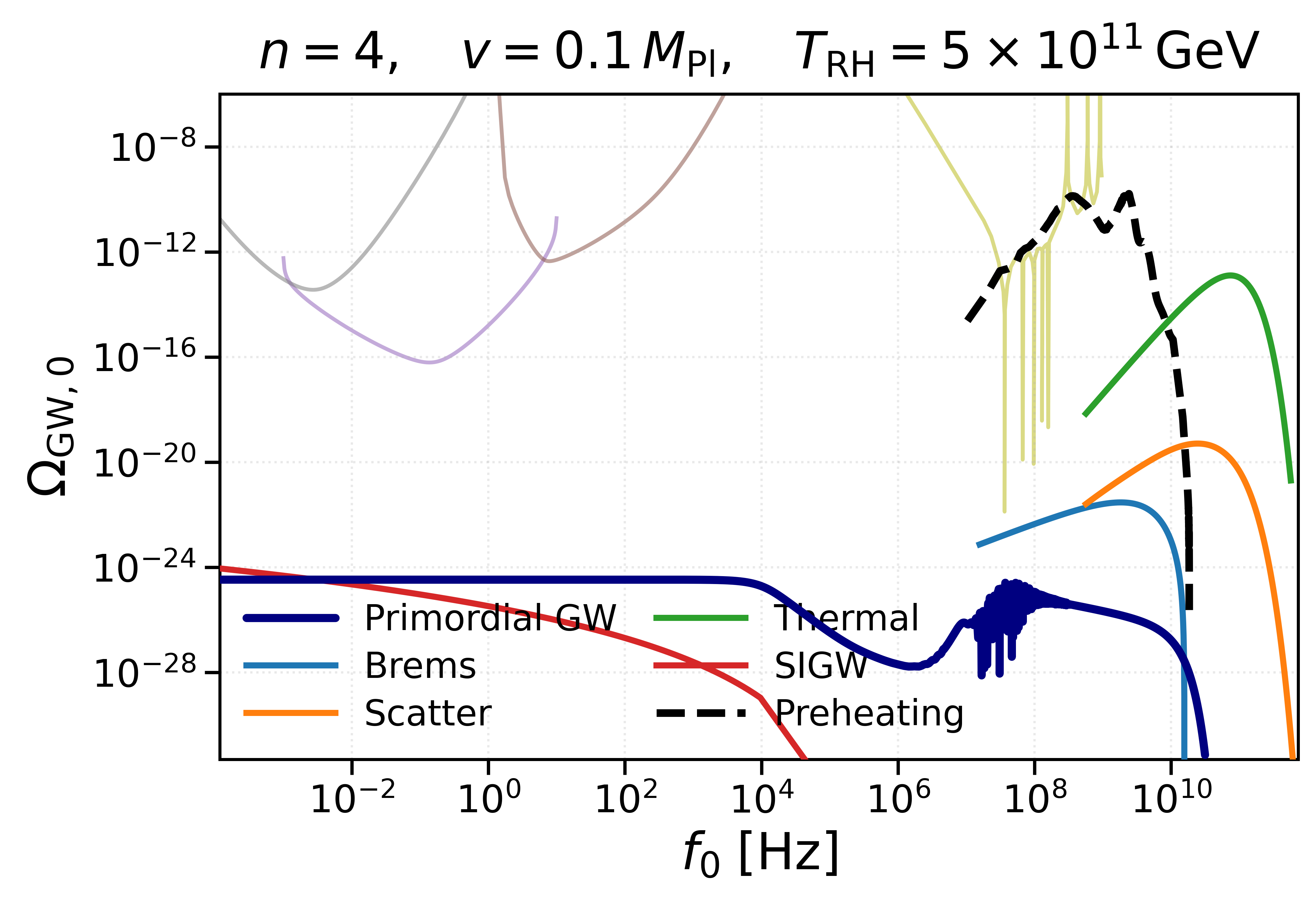}
   \includegraphics[width=0.48\textwidth]{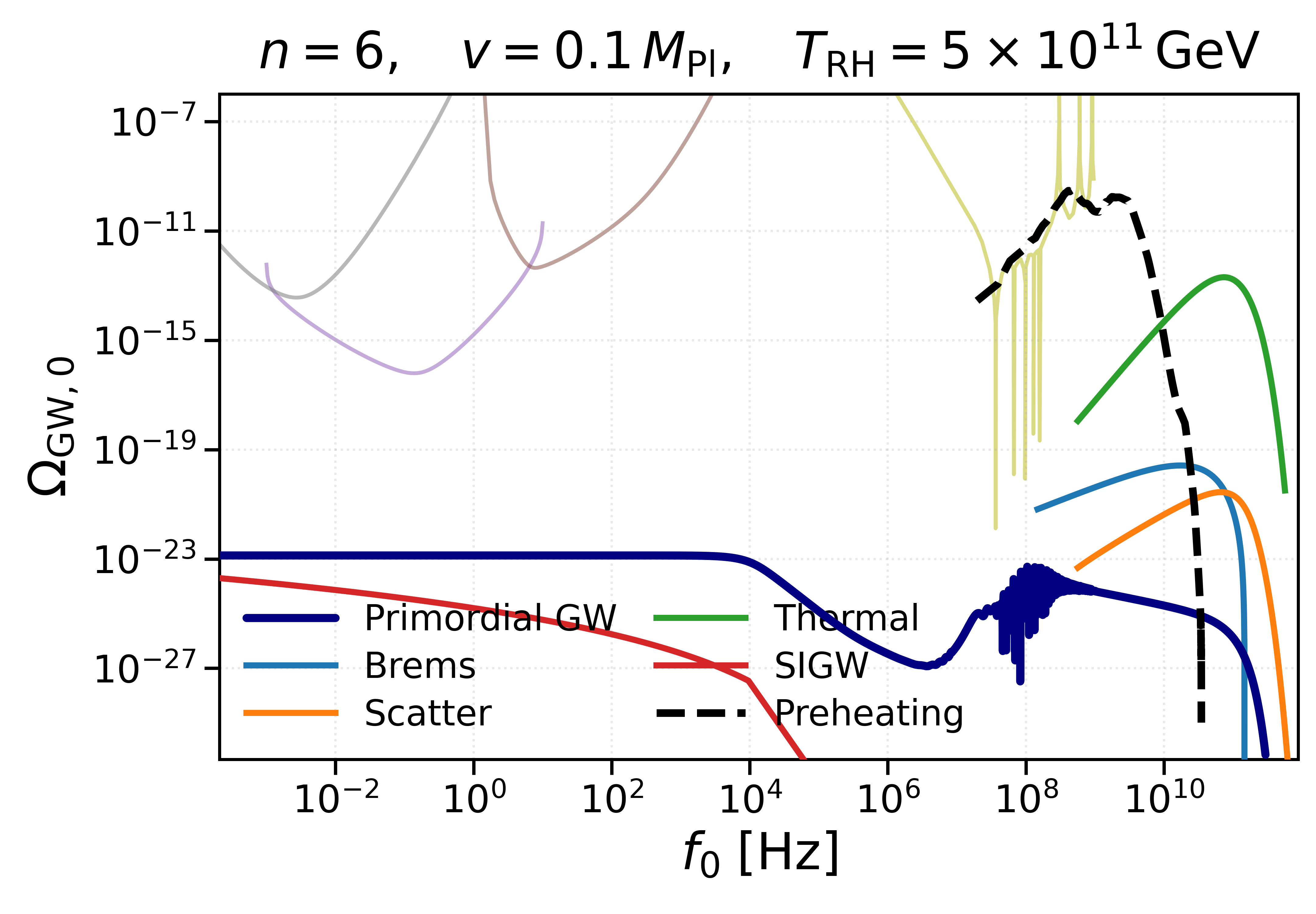}
   \includegraphics[width=0.48\textwidth]{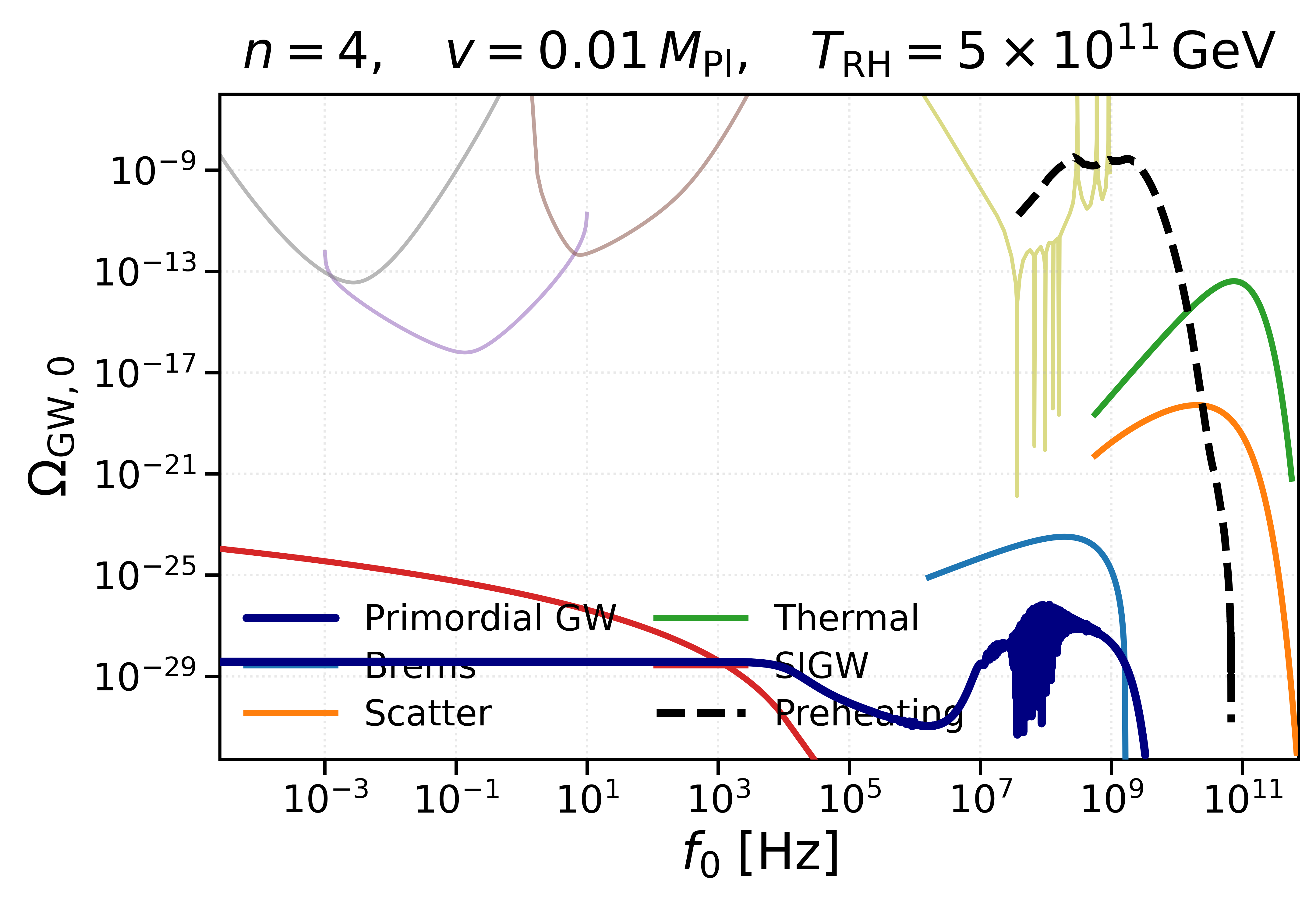}
   \includegraphics[width=0.48\textwidth]{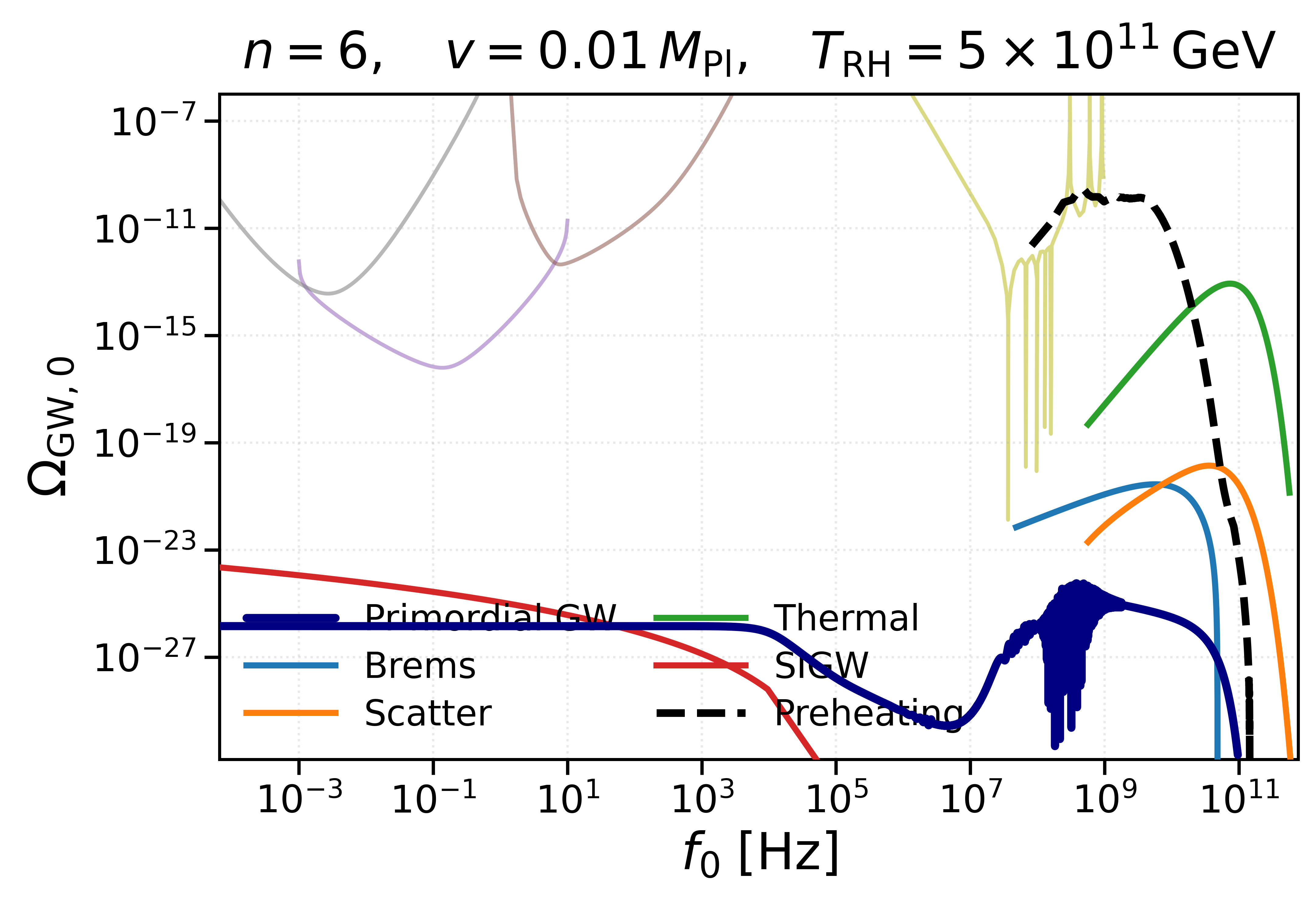}
  \end{center}
  \caption{
    Same as Fig.~\ref{fig:gw10}, but for $T_{\rm RH}=5\times 10^{11}\,{\rm GeV}$.}
  \label{fig:gw11}
\end{figure}

\begin{figure}
\begin{center}
   \includegraphics[width=0.48\textwidth]{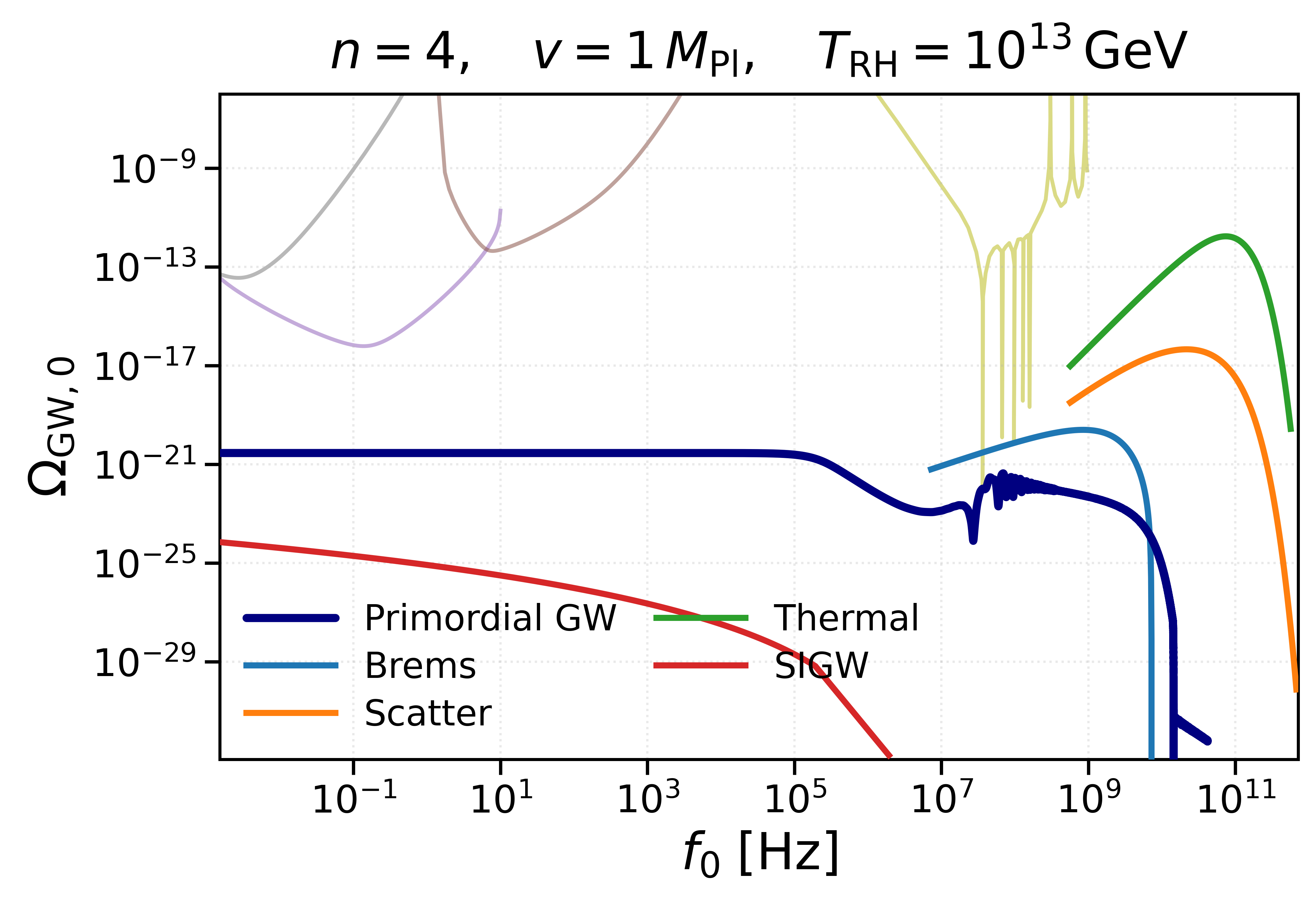}
   \includegraphics[width=0.48\textwidth]{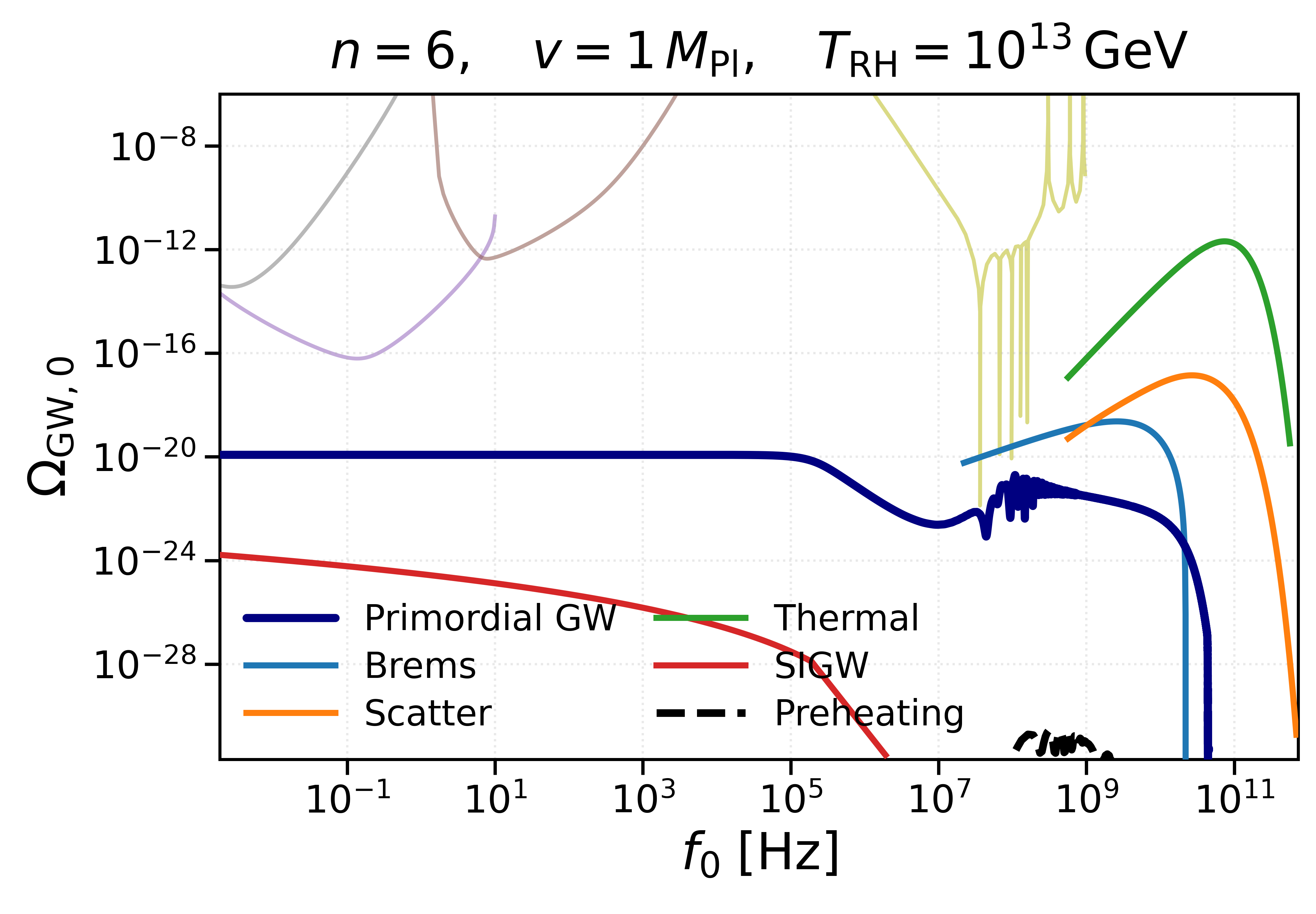}
   \includegraphics[width=0.48\textwidth]{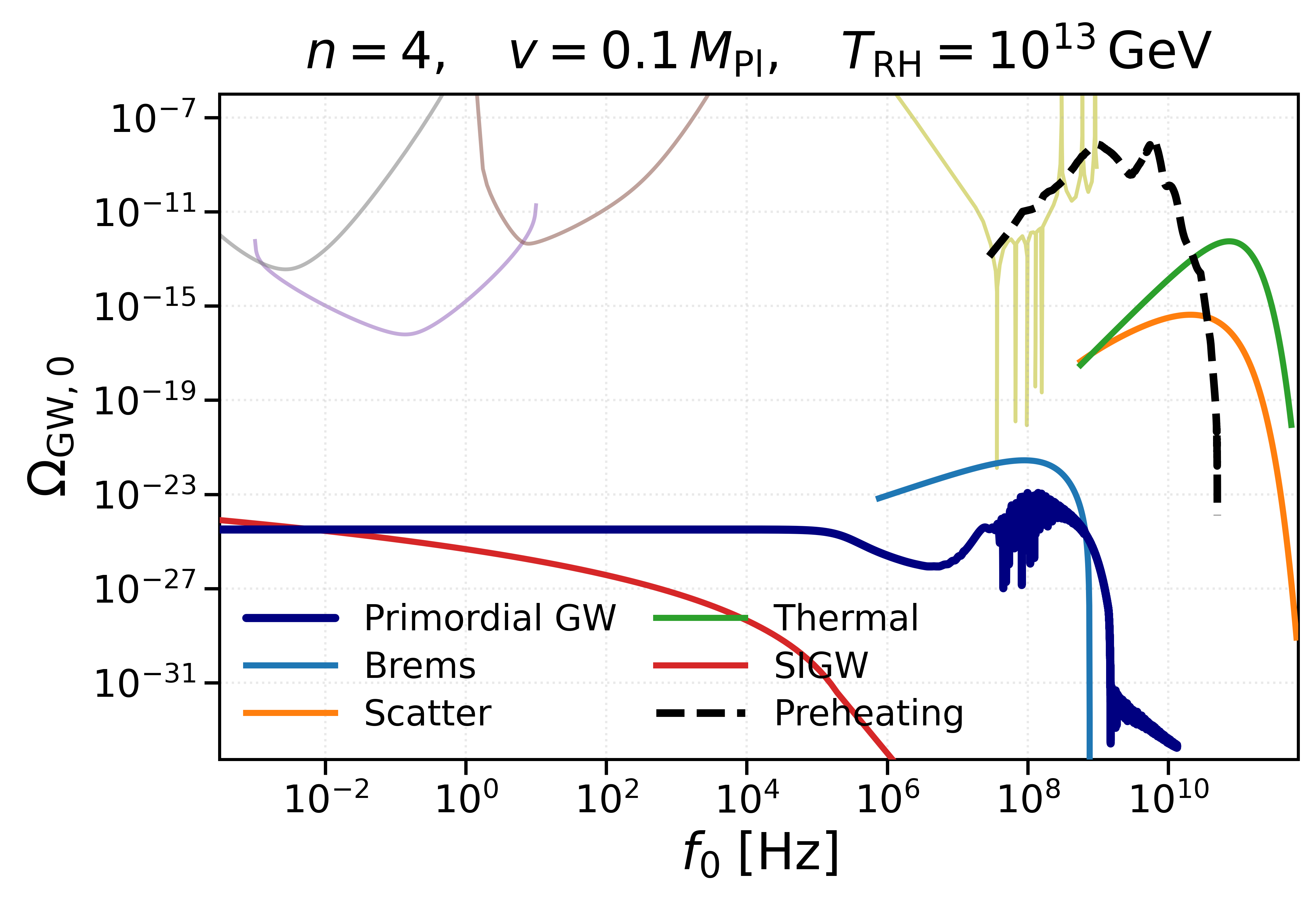}
   \includegraphics[width=0.48\textwidth]{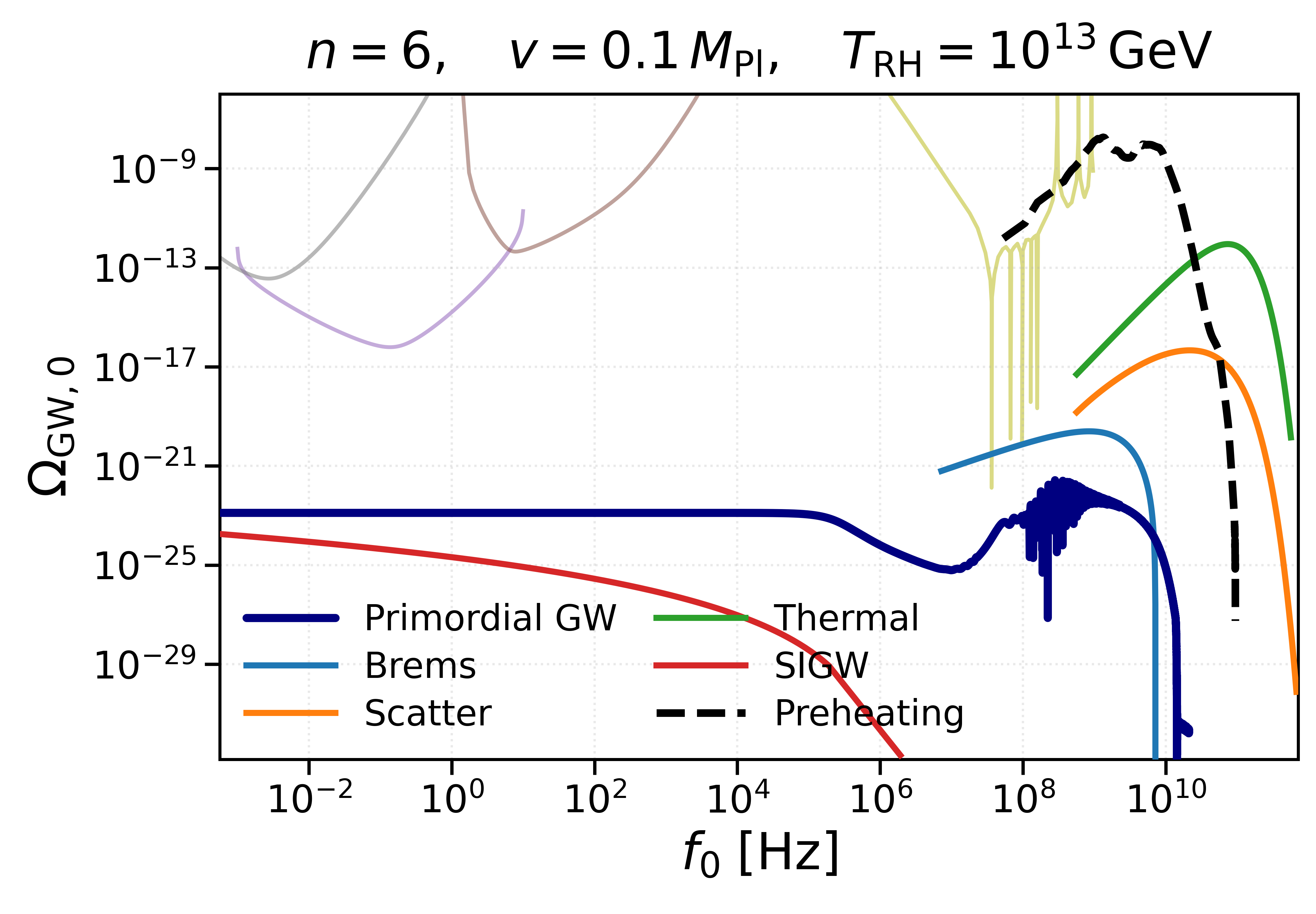}
   \includegraphics[width=0.48\textwidth]{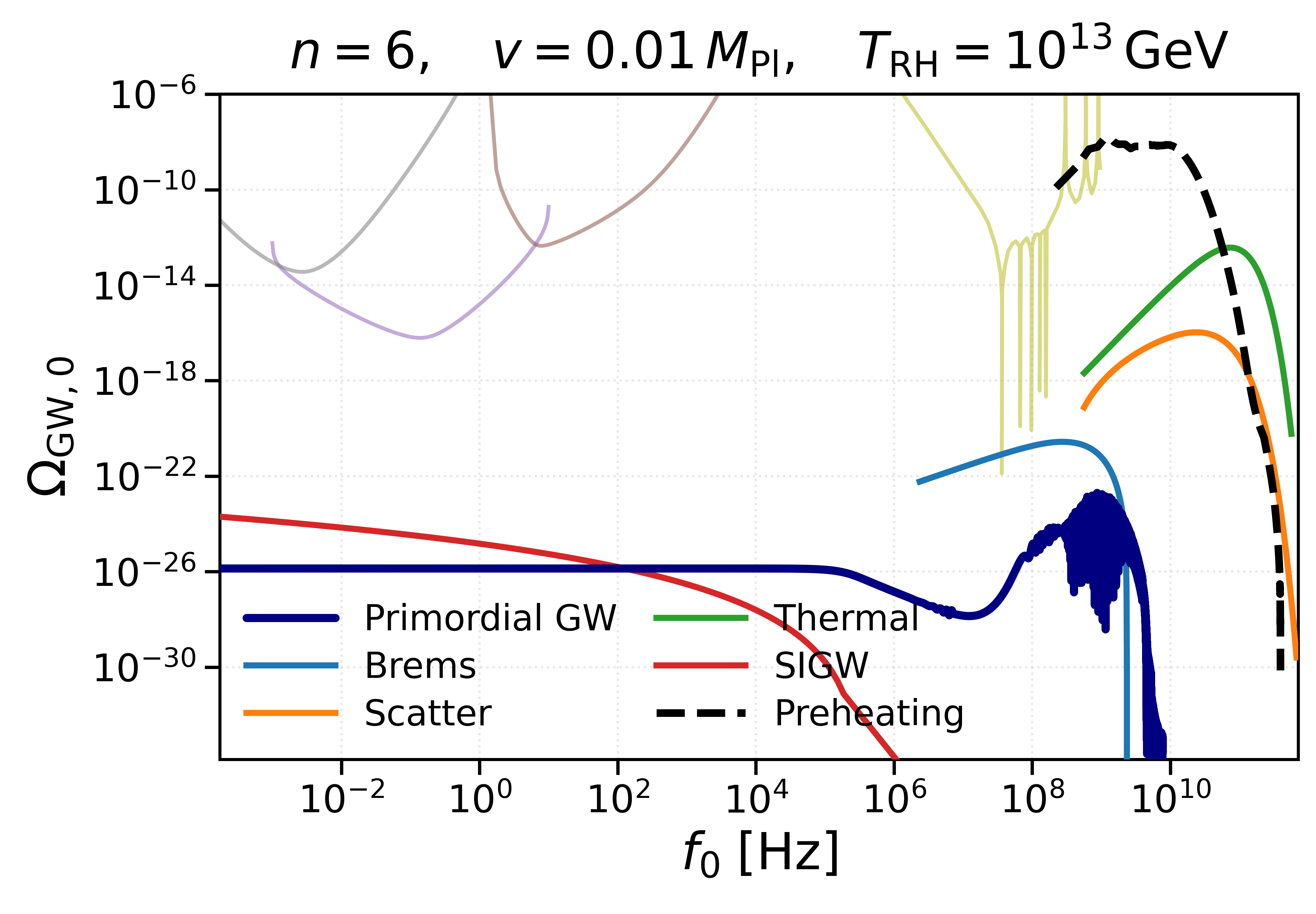}
  \end{center}
  \caption{
Same as Fig.~\ref{fig:gw10}, but for $T_{\rm RH}=10^{13}\,{\rm GeV}$.}
  \label{fig:gw13}
\end{figure}

Figs.~\ref{fig:gw10}-\ref{fig:gw13} show the resulting full GW spectra.
In Fig.~\ref{fig:gw10}, GW spectra with $n=4$ (left 3 plots) and $n=6$ (right 3 plots) for $T_{\rm RH}=10^{10}\,\rm GeV$ are shown. 
Displayed are the results for $v=M_{\rm Pl}$, $v=0.1M_{\rm Pl}$ and $v=0.01M_{\rm Pl}$, from top to bottom. 
Fig.~\ref{fig:gw11} and \ref{fig:gw13} are the same plots except for $T_{\rm RH}=5\times 10^{11}\,\rm GeV$ and $T_{\rm RH}=10^{13}\,\rm GeV$, respectively.
Sensitivities of current/future projects are also shown: SKA~\cite{Janssen:2014dka}, LISA~\cite{Bartolo:2016ami}, DECIGO~\cite{Yagi:2011wg}, ET~\cite{Punturo:2010zz},\footnote{
    For these lines we used a data file provided in Ref.~\cite{Schmitz:2020syl}.
} 
and resonant cavity~\cite{Herman:2020wao,Herman:2022fau}.

\section{Conclusions}
\label{sec:conc}

Precision measurements of CMB provided ways to constrain inflation models.
But still it is difficult to solve degeneracy among many inflation models by CMB or large scale cosmological measurements alone.
On the other hand, there have been recent theoretical developments on the production mechanisms of high frequency GWs from inflaton, which potentially contains rich information about inflation models.
In this paper we have presented full GW spectra in the new inflation model.
We performed lattice simulations to estimate GWs from preheating after inflation, and found that the preheating is efficient for $v\ll M_{\rm Pl}$ and hence huge amount of GWs are predicted.
We also calculated various quantum production processes of high frequency GWs, including inflaton annihilation, bremsstrahlung, and inflaton-particle scatterings.
They constitute GW background at even higher frequency than the preheating one with an amount depending on the model parameter.
This shows that the GW spectra at high frequency range can be used as a probe of inflation and its reheating dynamics.

Several comments are in order.
Although we presented GW spectra from various sources, still our calculations should not be regarded as robust and precise predictions. 
For example, the most calculations of quantum production assume the zero momentum of the inflaton, but actually spatial fluctuations develop during the reheating process. 
Although we expect that the final results do not change much since the inflaton remains non-relativistic even after the efficient preheating (as shown by our lattice simulations), further investigations may be required for more precise estimation.
Another caution is that, in the calculation of inflaton-particle scatterings, we assumed that the incoming particles have thermal distribution, but actually they should start from nonthermal distribution and approach to thermal one later.
Inclusions of full thermalization processes and resulting GW spectra remain to be explored.

\section*{Acknowledgment}

This work was supported by World Premier International Research Center Initiative (WPI), MEXT, Japan.
This work was also supported by JSPS KAKENHI (Grant Number 24K07010 [KN], 26K00695 [KN], 26H00403 [KN], 26K00695 [NK]).
This work used computational resources of Fugaku supercomputer, provided by RIKEN Center for Computational Sciences, through the HPCI System Research Project (Project ID: hp250177, hp260117). 

\appendix
\section{Some technical details}

\subsection{Precise calculation of primordial GW spectrum}
\label{app:precise}

In Sec.~\ref{subsec:hftail} we calculated the primordial GW spectra in both the low- and high-frequency limit.
The low frequency GWs correspond to those once exit the horizon during inflation, while the high frequency ones never experienced superhorizon regime and amplified during tne reheating era. 
The latter can be interpreted as inflaton pair annihilation into the graviton pair.
Although such diagrammatic calculations work in the deep inflaton-oscillating regime,  
this approach fails at the transient epoch from inflation to the reheating, at which the inflaton dynamics cannot be approximated by a periodic oscillation with monomial potential.
In such a regime we have to turn to numerical methods~\cite{Mudrunka:2026kgm,Wang:2026ule,Wang:2026pff}. 
We present an efficient method for evaluating the spectrum for a large number of modes at the same time,  which significantly improves our previous calculation speed~\cite{Mudrunka:2026kgm}.

Our starting point is the equation of motion of the graviton~(\ref{eq:gravitonOscillator}). Written in terms of conformal time and the variable $\widetilde h_{k,\lambda} \equiv a \frac{M_{\rm Pl}}{\sqrt{2}}h_{k,\lambda} $, the equation of motion is
\begin{align}
	\widetilde h_{k,\lambda}'' + \omega_k^2 \widetilde h_{k,\lambda} = 0,~~~~~~
	 \omega_k^2 = k^2 - \frac{a''}{a}=k^2-a^2(2H^2 + \dot H).
\end{align}
For modes that never exit the horizon we can employ an expansion using the WKB basis
\begin{align}
	\widetilde h_{k,\lambda}(\tau) = \alpha_k(\tau) v_k(\tau) + \beta_k(\tau) v_k^*(\tau),
\end{align}
where
\begin{align}
    v_k(\tau) \equiv  \frac{1}{\sqrt{2\omega_k(\tau)}} e^{-i\Omega_k},~~~~~~\Omega_k(\tau) = \int \omega_k(\tau) d\tau.
\end{align}
The $\alpha$ and $\beta$ used here are called the Bogoliubov coefficients. In terms of these the equation of motion becomes
\begin{align}
	\alpha'_k(\tau) = \frac{\omega_k'}{2\omega_k} \beta_k e^{2i\Omega_k},~~~~~~
	\beta'_k(\tau) = \frac{\omega_k'}{2\omega_k} \alpha_k e^{-2i\Omega_k}.
	\label{eq_Bogo}
\end{align}
$\beta$ measures the renormalized graviton energy density
\begin{align}
	a^4(\tau)\rho_h(\tau) &= 2 \int \frac{d^3k}{(2\pi)^3} \, k \, |\beta_k(\tau)|^2, ~~~	a^4(\tau) \rho_h(k, \tau) = \frac{k^4}{\pi^2} |\beta_k(\tau)|^2.
\end{align}
We have two coefficients $\alpha$ and $\beta$, but in reality only one of them is independent, because they are constructed to satisfy the normalization condition
\begin{equation}
    \vert \alpha_k\vert^2-\vert\beta_k\vert^2=1.
\end{equation}
We can use this to get rid of one of the equations. Since $\alpha_k\neq0$ at all times, we can define the variable 
\begin{equation}
    r_k=\frac{\beta_k}{\alpha_k},
\end{equation}
and get a single equation
\begin{equation}
    r_k^\prime=\frac{\omega_k^\prime}{2\omega_k}\left(e^{-2i\Omega_k}-r_k^2 e^{2i\Omega_k}\right).
\end{equation}
Integrating both sides, we get
\begin{equation}
    r_k=\int \frac{\omega_k^\prime}{2\omega_k}e^{-2i\Omega_k}d\tau-\int r_k^2\frac{\omega_k^\prime}{2\omega_k}e^{2i\Omega_k}d\tau.
\end{equation}
The first term corresponds to a first order time dependent perturbation theory. It describes the production of gravitons that do not further interact. The second term accounts for the higher order and nonperturbative effects which describe the backreaction of the already produced gravitons. As low as the graviton density is small $\vert r_k\vert^2\ll1$ the backreaction can be neglected and we are left with a fast oscillating integral to solve. 

For deeply subhorizon modes we have $k^2/a^2\gg2H^2+\dot{H}$, so $\Omega_k$ becomes just $k\tau$. Under this assumption we can write the integral as
\begin{equation}
    r_k=-\frac{1}{4k^2}\int U^\prime(\tau)e^{-2ik\tau}d\tau,~~~~~~U(\tau)=\frac{a^{\prime\prime}}{a}.
\end{equation}
The resulting spectrum is nothing but a Fourier transform of $U^\prime$. This can be computed very efficiently for all $k$ at once using any existing fast-Fourier-transform (FFT) algorithm, making the background evolution the bottleneck. Furthermore we can avoid having to compute $U^\prime$ by integrating by parts
\begin{equation}
    r_k=-\frac{1}{4k^2}\left[U(\tau)e^{-2ik\tau}\right]_{\tau_0}^{\tau_{\rm end}}-\frac{i}{2k}\int U(\tau)e^{-2ik\tau}d\tau.
\end{equation}
With this method, we can numerically evaluate $r_k$ very efficiently and hence $\beta_k$ (under an assumption $|\beta_k|\ll 1$) and $\rho_h(k)$ accordingly.  
It is much faster than solving the graviton equation of motion or the equation for $\beta_k$ (\ref{eq_Bogo}) directly.

\subsection{Collision term for inflaton-particle scatter}
\label{appendix:scatter}
In this section we present the detailed calculation of the graviton production rate through $\phi\chi\rightarrow h\chi$ scattering used in section \ref{sec:scatter}. 

\paragraph{Kinematics}

First let us see the kinematics of this scattering process.
Let us suppose that $\chi$ is massless and the inflaton is at rest, i.e., $l_\mu=(m_\phi, \vec 0)$. 
Without loss of generality, we can take the incoming particle momentum as $p_\mu=(p,p,0,0)$.
By taking the graviton four momentum as $k_\mu=(E_k,E_k\cos\theta,E_k\sin\theta,0)$ and the out-going $\chi$ momentum as $q_\mu=(q, q\cos\tilde\theta,-q\sin\tilde\theta,0)$, and using the energy-momentum conservation
\begin{equation}
\label{eq:scatteringMomentaRelation}
    l_\mu+p_\mu=q_\mu+k_\mu,
\end{equation}
we obtain
\begin{equation}
   E_k = \frac{m_\phi}{2} \frac{m_\phi + 2p}{m_\phi + p(1-\cos\theta)},
   \label{Ek_theta}
\end{equation}
or equivalently,
\begin{equation}
    p= \frac{m_\phi}{2} \frac{2E_k-m_\phi}{E_k(\cos\theta-1)+m_\phi}~~(\equiv p_*).
    \label{p*}
\end{equation}
From this expression, we can deduce several things. First, $E_k \geq m_\phi/2$ always holds.
In the heavy inflaton limit $m_\phi\gg p$, we have $E_k \simeq m_\phi/2$ independently of the scattering angle $\theta$. This may be interpreted as the inflaton decay almost at rest, slightly assisted by an incoming particle.
In the opposite case $m_\phi\ll E_k$, energy of the final state graviton is peaked around $\theta \lesssim \sqrt{2m_\phi/p}\equiv \theta_*$ at which $E_k \simeq p$. Otherwise it shows angular dependence like $E_k \simeq m_\phi/(1-\cos\theta)$ for $\theta\gtrsim \theta_*$.
In general, Eq.~(\ref{p*}) and the condition $p>0$ restrict the angle as $\cos\theta > 1-m_\phi/E_k$ for given $E_k$.

\paragraph{Matrix element}

The squared scattering amplitude for one real scalar $\chi$ is obtained as
\begin{equation}
    \vert \mathcal  M\vert^2=\sum_{+\times}\frac{\lambda^2\kappa^2}{4}\left(\frac{q_\mu q_\nu}{q\cdot k}-\frac{p_\mu p_\nu}{p\cdot k}\right)\left(\frac{q_\alpha q_\beta}{q\cdot k}-\frac{p_\alpha p_\beta}{p\cdot k}\right)\epsilon^{\mu\nu}\epsilon^{*\alpha\beta}.
\end{equation} 
Here and hereafter we define $\kappa \equiv 2/M_{\rm Pl}$.
Using \eqref{eq:polarizationInflaton} and \eqref{eq:scatteringMomentaRelation}, we can replace all $q_{\mu}$ in the numerators by $p_{\mu}$ inside the polarization sum, so that we obtain
\begin{equation}
    \vert \mathcal M\vert^2=\sum_{+\times}\frac{\lambda^2\kappa^2}{4}\left[\frac{(p-q)\cdot k}{(p\cdot k)(q \cdot k)}\right]^2p_{\mu}p_{\nu}p_{\alpha}p_{\beta}\epsilon^{\mu\nu}\epsilon^{*\alpha\beta}.
\end{equation}
Noting that the graviton polarization tensors satisfy the following identity
\begin{equation}
    \sum_{+\times}\epsilon^{*\mu\nu}\epsilon^{\alpha\beta}=\frac{1}{2}\left(\hat{\eta}^{\mu\alpha
    }\hat{\eta}^{\nu\beta}+\hat{\eta}^{\mu\beta}\hat{\eta}^{\nu\alpha}-\hat{\eta}^{\mu\nu}\hat{\eta}^{\alpha\beta}\right),~~~\hat{\eta}_{\mu\nu}=\eta_{\mu\nu}-\frac{k_{\mu}\bar{k}_{\nu}+\bar{k}_{\mu}k_{\nu}}{2E_{k}^2},
\end{equation}
we have
\begin{equation}
    \sum_{+\times}p_{\mu}p_{\nu}p_{\alpha}p_{\beta}\epsilon^{\mu\nu}\epsilon^{*\alpha\beta}=\frac{1}{2}\left[(p\cdot p)-\frac{(p\cdot k)(p\cdot \bar{k})}{E_k^2}\right]^2.
\end{equation}
Approximating $\chi$ as massless, this simplifies to
\begin{equation}
    \frac{1}{2}\left[(p\cdot p)-\frac{(p\cdot k)(p\cdot \bar{k})}{E_k^2}\right]^2=\frac{1}{2}p^4\sin^4\theta,
\end{equation}
where $\theta$ is the angle between $\vec p$ and $\vec k$. The first term evaluates to
\begin{equation}
    \left[\frac{(p-q)\cdot k}{(p\cdot k)(q \cdot k)}\right]^2=\frac{m_{\phi}^2}{E_k^2 p^2\left(1-\cos\theta\right)^2\left[p\left(1-\cos\theta\right)+m_{\phi}\right]^2}.
\end{equation}
Combining these, we get
\begin{align}
    \vert \mathcal M\vert^2 =\frac{\lambda^2\kappa^2m_{\phi}^2 p^2\sin^4\theta}{8E_k^2 \left(1-\cos\theta\right)^2\left[p\left(1-\cos\theta\right)+m_{\phi}\right]^2}.
    \label{scatter_M2}
\end{align}
By using \eqref{p*}, we find that the square amplitude (\ref{scatter_M2}) is rewritten as
\begin{align}
    \left|\mathcal M\right|^2=\frac{\lambda^2\kappa^2}{8}\left(2-\frac{m_\phi}{E_k}\right)^2,
\end{align}
which is consistent with Ref.~\cite{Xu:2024fjl,Bernal:2025lxp}.

\paragraph{Boltzmann equation}

The Boltzmann equation for graviton phase space distribution $f_h(t,E_k)$ is expressed as
\begin{equation}
    \frac{\partial f_h}{\partial t} - HE_k \frac{\partial f_h}{\partial E_k} = C,
    \label{f_Boltzmann}
\end{equation}
where the collision term is
\begin{equation}
    C=\frac{1}{2E_k}\int\frac{d^3 q}{(2\pi)^3 2q}\frac{d^3 p}{(2\pi)^3 2p}\frac{d^3 l}{(2\pi)^3 2m_{\phi}}(2\pi)^4 \delta^{(4)}(l_\mu+p_\mu-q_\mu-k_\mu)\vert \mathcal M\vert^2\frac{f_{\phi}}{e^{p/T}-1}.
\end{equation}
The last term comes from the assumption that $\chi$ is thermalized and follows the Bose-Einstein distribution. 
The inflaton is by design considered to be at rest and we can take $f_{\phi}=(2\pi)^3\rho_{\phi}/m_{\phi}\delta^{(3)}(\vec l)$, so the integration over $d^3 l$ is trivial and we get
\begin{equation}
    C=\frac{1}{2E_{k}}\frac{\rho_{\phi}}{8(2\pi)^2m_{\phi}^2}\int \frac{d^3q d^3p}{qp}\delta^{(4)}(l_\mu+p_\mu-q_\mu-k_\mu)\frac{\vert \mathcal M\vert ^2}{e^{p/T}-1}.
\end{equation}
Integrating over $d^3q$ removes the spatial part of the delta function and gives us $q=p-k$, and so 
\begin{equation}
\label{eq:momentumDeltaIntegrationResult}
    \frac{1}{2E_{k}}\frac{\rho_{\phi}}{8(2\pi)^2m_{\phi}^2}\int \frac{d^3p}{qp}\delta^{(0)}(m_\phi+p-q-E_k)\frac{\vert \mathcal M\vert ^2}{e^{p/T}-1},~~~q^2=p^2-2pE_k\cos\theta+E_k^2.
\end{equation}
Going to spherical coordinates $d^3p=d\Omega_p\, p^2dp$, this becomes
\begin{equation}
    \frac{1}{2E_{k}}\frac{\rho_{\phi}}{8(2\pi)^2m_{\phi}^2}\int \frac{d\Omega p^2dp}{qp}\delta^{(0)}(m_{\phi}+p-q-E_k)\frac{\vert \mathcal M\vert ^2}{e^{p/T}-1}.
\end{equation}
Using
\begin{equation}
    \delta(f(x))=\sum_{i}\frac{\delta(x-x_i)}{\vert f^{\prime}(x_i) \vert},
\end{equation}
and
\begin{equation}
    \frac{d(p-q)}{dp}=\frac{q-p+E_k\cos\theta}{q},
\end{equation}
we obtain
\begin{equation}
    C=\frac{1}{2E_{k}}\frac{\rho_{\phi}}{8(2\pi)^2m_{\phi}^2}\int d\Omega_p \frac{p_*\vert \mathcal M \vert^2}{q(p_*)-p_*+E_k\cos\theta}\frac{1}{e^{p_*/T}-1},
\end{equation}
where $p_*$ is the root of $0= m_{\phi}+p-q-E_k$ as function of $p$, given by (\ref{p*}). Equivalently, $E_k$ is expressed as (\ref{Ek_theta}).
Then the collision term becomes
\begin{equation}
    C=\frac{\kappa^{2} \lambda^{2} }{256E_{k}^3 m_\phi}\frac{\rho_{\phi}}{(2\pi)^2}\int d\Omega_p\frac{(2E_k-m_\phi)^3}{\left(E_{k} \cos{\theta} - E_{k} + m_{\phi}\right)^{2}}\frac{1}{e^{\frac{m_{\phi} \left(2 E_{k} - m_{\phi}\right)}{2 \left(E_{k} \cos{\theta} - E_{k} + m_{\phi}\right)T}}-1}.
\end{equation}
Integrating over the azimuthal angle and substituting for a new variable $Z=E_k\cos\theta-E_k+m_{\phi}$, by noting that $dZ=-E_k\sin(\theta)d\theta$, we arrive at
\begin{equation}
    C=\frac{\rho_{\phi}}{m_{\phi}}\frac{\kappa^{2} \lambda^{2} }{256E_{k}^4}\frac{(2E_k-m_\phi)^3}{2\pi}\int^{m_{\phi}}_{0} \frac{dZ}{Z^2\left(e^{\frac{m_{\phi}(2E_k-m_{\phi})}{2ZT}}-1\right)}.
\end{equation}
The integral can be easily evaluated by another substitution and we finally get
\begin{align}
    C&=\frac{\rho_{\phi}}{m_{\phi}}\frac{\kappa^{2} \lambda^{2} }{256E_{k}^4}\frac{(2E_k-m_\phi)^3}{2\pi}\frac{2 E_{k} - 2 T \ln{\left(e^{\frac{2 E_{k} - m_{\phi}}{2 T}} - 1 \right)} - m_{\phi}}{m_{\phi} \left(2 E_{k} - m_{\phi}\right)}\\
    &=\frac{\rho_{\phi}}{m_{\phi}}\frac{\kappa^{2} \lambda^{2} }{512\pi E_{k}^4 m_{\phi}}\left(2E_k-m_{\phi}\right)^3\left[1-\frac{\ln(e^x-1)}{x}\right],~~~~~~x=\frac{2E_k-m_{\phi}}{2T}.
\end{align}

From Eq.~(\ref{f_Boltzmann}), the Boltzmann equation for graviton energy spectrum $d\rho_h/dE_k = f_hE_k^3/(2\pi^2)$ is given by
\begin{equation}
    \left[\frac{\partial}{\partial t} +3H- HE_k \frac{\partial}{\partial E_k}\right] \left(\frac{d\rho_h}{dE_k}\right)= \frac{d\Gamma}{dE_k},
    \label{rho_Boltzmann}
\end{equation}
where
\begin{equation}
    \frac{d\Gamma}{dE_k}=\frac{\rho_{\phi}}{m_{\phi}^2}\frac{\kappa^{2} \lambda^{2} }{256\pi^3 E_k}\left(2E_k-m_{\phi}\right)^3\left[1-\frac{\ln(e^x-1)}{x}\right],~~~x=\frac{2E_k-m_{\phi}}{2T}.
\end{equation}
Here we have taken into account the four real degrees of freedom of the Higgs.

\bibliography{ref}

@article{Planck:2018jri,
    author = "Akrami, Y. and others",
    collaboration = "Planck",
    title = "{Planck 2018 results. X. Constraints on inflation}",
    eprint = "1807.06211",
    archivePrefix = "arXiv",
    primaryClass = "astro-ph.CO",
    doi = "10.1051/0004-6361/201833887",
    journal = "Astron. Astrophys.",
    volume = "641",
    pages = "A10",
    year = "2020"
}

@article{Planck:2018vyg,
    author = "Aghanim, N. and others",
    collaboration = "Planck",
    title = "{Planck 2018 results. VI. Cosmological parameters}",
    eprint = "1807.06209",
    archivePrefix = "arXiv",
    primaryClass = "astro-ph.CO",
    doi = "10.1051/0004-6361/201833910",
    journal = "Astron. Astrophys.",
    volume = "641",
    pages = "A6",
    year = "2020",
    note = "[Erratum: Astron.Astrophys. 652, C4 (2021)]"
}

@article{Martin:2013tda,
    author = "Martin, Jerome and Ringeval, Christophe and Vennin, Vincent",
    title = "{Encyclop{\ae}dia Inflationaris}: {Opiparous Edition}",
    eprint = "1303.3787",
    archivePrefix = "arXiv",
    primaryClass = "astro-ph.CO",
    doi = "10.1016/j.dark.2024.101653",
    journal = "Phys. Dark Univ.",
    volume = "5-6",
    pages = "75--235",
    year = "2014"
}

@article{Starobinsky:1979ty,
    author = "Starobinsky, Alexei A.",
    editor = "Khalatnikov, I. M. and Mineev, V. P.",
    title = "{Spectrum of relict gravitational radiation and the early state of the universe}",
    journal = "JETP Lett.",
    volume = "30",
    pages = "682--685",
    year = "1979"
}

@article{Linde:1981mu,
    author = "Linde, Andrei D.",
    editor = "Fang, Li-Zhi and Ruffini, R.",
    title = "{A New Inflationary Universe Scenario: A Possible Solution of the Horizon, Flatness, Homogeneity, Isotropy and Primordial Monopole Problems}",
    reportNumber = "LEBEDEV-81-229",
    doi = "10.1016/0370-2693(82)91219-9",
    journal = "Phys. Lett. B",
    volume = "108",
    pages = "389--393",
    year = "1982"
}

@article{Albrecht:1982wi,
    author = "Albrecht, Andreas and Steinhardt, Paul J.",
    editor = "Fang, Li-Zhi and Ruffini, R.",
    title = "{Cosmology for Grand Unified Theories with Radiatively Induced Symmetry Breaking}",
    reportNumber = "UPR-0185T",
    doi = "10.1103/PhysRevLett.48.1220",
    journal = "Phys. Rev. Lett.",
    volume = "48",
    pages = "1220--1223",
    year = "1982"
}

@book{Liddle:2000cg,
    author = "Liddle, Andrew R. and Lyth, D. H.",
    title = "{Cosmological inflation and large scale structure}",
    doi = "10.1017/CBO9781139175180",
    isbn = "978-0-521-57598-0, 978-0-521-82849-9",
    year = "2000"
}

@article{Kumekawa:1994gx,
    author = "Kumekawa, Kazuya and Moroi, Takeo and Yanagida, Tsutomu",
    title = "{Flat potential for inflaton with a discrete R invariance in supergravity}",
    eprint = "hep-ph/9405337",
    archivePrefix = "arXiv",
    reportNumber = "TU-458",
    doi = "10.1143/PTP.92.437",
    journal = "Prog. Theor. Phys.",
    volume = "92",
    pages = "437--448",
    year = "1994"
}

@article{Izawa:1996dv,
    author = "Izawa, K. -I. and Yanagida, T.",
    title = "{Natural new inflation in broken supergravity}",
    eprint = "hep-ph/9608359",
    archivePrefix = "arXiv",
    reportNumber = "UT-757",
    doi = "10.1016/S0370-2693(96)01638-3",
    journal = "Phys. Lett. B",
    volume = "393",
    pages = "331--336",
    year = "1997"
}

@article{Izawa:1997df,
    author = "Izawa, K. I. and Kawasaki, M. and Yanagida, T.",
    title = "{Dynamical tuning of the initial condition for new inflation in supergravity}",
    eprint = "hep-ph/9707201",
    archivePrefix = "arXiv",
    reportNumber = "UT-780, ICRR-396-97-19",
    doi = "10.1016/S0370-2693(97)01040-X",
    journal = "Phys. Lett. B",
    volume = "411",
    pages = "249--255",
    year = "1997"
}

@article{Asaka:1999jb,
    author = "Asaka, T. and Hamaguchi, Koichi and Kawasaki, M. and Yanagida, T.",
    title = "{Leptogenesis in inflationary universe}",
    eprint = "hep-ph/9907559",
    archivePrefix = "arXiv",
    reportNumber = "UT-855, RESCEU-28-99",
    doi = "10.1103/PhysRevD.61.083512",
    journal = "Phys. Rev. D",
    volume = "61",
    pages = "083512",
    year = "2000"
}

@article{Senoguz:2004ky,
    author = "Senoguz, Vedat Nefer and Shafi, Q.",
    title = "{New inflation, preinflation, and leptogenesis}",
    eprint = "hep-ph/0403294",
    archivePrefix = "arXiv",
    reportNumber = "BA-04-04",
    doi = "10.1016/j.physletb.2004.05.077",
    journal = "Phys. Lett. B",
    volume = "596",
    pages = "8--15",
    year = "2004"
}

@article{Kohri:2007gq,
    author = "Kohri, Kazunori and Lin, Chia-Min and Lyth, David H.",
    title = "{More hilltop inflation models}",
    eprint = "0707.3826",
    archivePrefix = "arXiv",
    primaryClass = "hep-ph",
    doi = "10.1088/1475-7516/2007/12/004",
    journal = "JCAP",
    volume = "12",
    pages = "004",
    year = "2007"
}

@article{Nakayama:2012dw,
    author = "Nakayama, Kazunori and Takahashi, Fuminobu",
    title = "{PeV-scale Supersymmetry from New Inflation}",
    eprint = "1203.0323",
    archivePrefix = "arXiv",
    primaryClass = "hep-ph",
    reportNumber = "UT-12-07, TU-902, IPMU12-0028",
    doi = "10.1088/1475-7516/2012/05/035",
    journal = "JCAP",
    volume = "05",
    pages = "035",
    year = "2012"
}

@article{Takahashi:2013cxa,
    author = "Takahashi, Fuminobu",
    title = "{New inflation in supergravity after Planck and LHC}",
    eprint = "1308.4212",
    archivePrefix = "arXiv",
    primaryClass = "hep-ph",
    reportNumber = "TU-943, IPMU13-0159",
    doi = "10.1016/j.physletb.2013.10.026",
    journal = "Phys. Lett. B",
    volume = "727",
    pages = "21--26",
    year = "2013"
}

@article{Ema:2017rkk,
    author = "Ema, Yohei and Mukaida, Kyohei and Nakayama, Kazunori",
    title = "{Electroweak Vacuum Metastability and Low-scale Inflation}",
    eprint = "1706.08920",
    archivePrefix = "arXiv",
    primaryClass = "hep-ph",
    reportNumber = "UT-17-21, IPMU-17-0090",
    doi = "10.1088/1475-7516/2017/12/030",
    journal = "JCAP",
    volume = "12",
    pages = "030",
    year = "2017"
}

@article{Dimopoulos:2020kol,
    author = "Dimopoulos, Konstantinos",
    title = "{An analytic treatment of quartic hilltop inflation}",
    eprint = "2006.06029",
    archivePrefix = "arXiv",
    primaryClass = "hep-ph",
    doi = "10.1016/j.physletb.2020.135688",
    journal = "Phys. Lett. B",
    volume = "809",
    pages = "135688",
    year = "2020"
}

@article{Dolgov:1989us,
    author = "Dolgov, A. D. and Kirilova, D. P.",
    title = "{ON PARTICLE CREATION BY A TIME DEPENDENT SCALAR FIELD}",
    reportNumber = "JINR-E2-89-321",
    journal = "Sov. J. Nucl. Phys.",
    volume = "51",
    pages = "172--177",
    year = "1990"
}

@article{Traschen:1990sw,
    author = "Traschen, Jennie H. and Brandenberger, Robert H.",
    title = "{Particle Production During Out-of-equilibrium Phase Transitions}",
    reportNumber = "BROWN-HET-731",
    doi = "10.1103/PhysRevD.42.2491",
    journal = "Phys. Rev. D",
    volume = "42",
    pages = "2491--2504",
    year = "1990"
}

@article{Shtanov:1994ce,
    author = "Shtanov, Y. and Traschen, Jennie H. and Brandenberger, Robert H.",
    title = "{Universe reheating after inflation}",
    eprint = "hep-ph/9407247",
    archivePrefix = "arXiv",
    reportNumber = "BROWN-HET-957",
    doi = "10.1103/PhysRevD.51.5438",
    journal = "Phys. Rev. D",
    volume = "51",
    pages = "5438--5455",
    year = "1995"
}

@article{Kofman:1994rk,
    author = "Kofman, Lev and Linde, Andrei D. and Starobinsky, Alexei A.",
    title = "{Reheating after inflation}",
    eprint = "hep-th/9405187",
    archivePrefix = "arXiv",
    reportNumber = "UH-IFA-94-35, SU-ITP-94-13, YITP-U-94-15",
    doi = "10.1103/PhysRevLett.73.3195",
    journal = "Phys. Rev. Lett.",
    volume = "73",
    pages = "3195--3198",
    year = "1994"
}

@article{Kofman:1997yn,
    author = "Kofman, Lev and Linde, Andrei D. and Starobinsky, Alexei A.",
    title = "{Towards the theory of reheating after inflation}",
    eprint = "hep-ph/9704452",
    archivePrefix = "arXiv",
    reportNumber = "IFA-97-28, SU-ITP-97-18",
    doi = "10.1103/PhysRevD.56.3258",
    journal = "Phys. Rev. D",
    volume = "56",
    pages = "3258--3295",
    year = "1997"
}

@article{Felder:2001kt,
    author = "Felder, Gary N. and Kofman, Lev and Linde, Andrei D.",
    title = "{Tachyonic instability and dynamics of spontaneous symmetry breaking}",
    eprint = "hep-th/0106179",
    archivePrefix = "arXiv",
    reportNumber = "CITA-2001-12, SU-ITP-01-26",
    doi = "10.1103/PhysRevD.64.123517",
    journal = "Phys. Rev. D",
    volume = "64",
    pages = "123517",
    year = "2001"
}

@article{Felder:2000hj,
    author = "Felder, Gary N. and Garcia-Bellido, Juan and Greene, Patrick B. and Kofman, Lev and Linde, Andrei D. and Tkachev, Igor",
    title = "{Dynamics of symmetry breaking and tachyonic preheating}",
    eprint = "hep-ph/0012142",
    archivePrefix = "arXiv",
    reportNumber = "CITA-2000-60, SU-ITP-00-35, IFT-UAM-CSIC-00-40, FT-UAM-00-26, CERN-TH-2000-365",
    doi = "10.1103/PhysRevLett.87.011601",
    journal = "Phys. Rev. Lett.",
    volume = "87",
    pages = "011601",
    year = "2001"
}

@article{Desroche:2005yt,
    author = "Desroche, Mariel and Felder, Gary N. and Kratochvil, Jan M. and Linde, Andrei D.",
    title = "{Preheating in new inflation}",
    eprint = "hep-th/0501080",
    archivePrefix = "arXiv",
    reportNumber = "SU-ITP-5-02",
    doi = "10.1103/PhysRevD.71.103516",
    journal = "Phys. Rev. D",
    volume = "71",
    pages = "103516",
    year = "2005"
}

@article{Brax:2010ai,
    author = "Brax, Philippe and Dufaux, Jean-Francois and Mariadassou, Sophie",
    title = "{Preheating after Small-Field Inflation}",
    eprint = "1012.4656",
    archivePrefix = "arXiv",
    primaryClass = "hep-th",
    doi = "10.1103/PhysRevD.83.103510",
    journal = "Phys. Rev. D",
    volume = "83",
    pages = "103510",
    year = "2011"
}

@article{Antusch:2015nla,
    author = "Antusch, Stefan and Nolde, David and Orani, Stefano",
    title = "{Hill crossing during preheating after hilltop inflation}",
    eprint = "1503.06075",
    archivePrefix = "arXiv",
    primaryClass = "hep-ph",
    doi = "10.1088/1475-7516/2015/06/009",
    journal = "JCAP",
    volume = "06",
    pages = "009",
    year = "2015"
}

@article{Antusch:2015vna,
    author = "Antusch, Stefan and Cefala, Francesco and Nolde, David and Orani, Stefano",
    title = "{Parametric resonance after hilltop inflation caused by an inhomogeneous inflaton field}",
    eprint = "1510.04856",
    archivePrefix = "arXiv",
    primaryClass = "hep-ph",
    doi = "10.1088/1475-7516/2016/02/044",
    journal = "JCAP",
    volume = "02",
    pages = "044",
    year = "2016"
}

@article{Das:2025tzs,
    author = "Das, Diganta and Revankar, Shreyas",
    title = "{Preheating and gravitational waves in large-field hilltop inflation}",
    eprint = "2508.07442",
    archivePrefix = "arXiv",
    primaryClass = "astro-ph.CO",
    doi = "10.1140/epjs/s11734-025-01863-x",
    month = "8",
    year = "2025"
}

@article{Khlebnikov:1997di,
	archiveprefix = {arXiv},
	author = {Khlebnikov, S. Y. and Tkachev, I. I.},
	doi = {10.1103/PhysRevD.56.653},
	eprint = {hep-ph/9701423},
	journal = {Phys. Rev. D},
	pages = {653--660},
	reportnumber = {PURD-TH-97-02, OSU-TA-01-97},
	title = {{Relic gravitational waves produced after preheating}},
	volume = {56},
	year = {1997}}

@article{Easther:2006vd,
	archiveprefix = {arXiv},
	author = {Easther, Richard and Giblin, Jr., John T. and Lim, Eugene A.},
	doi = {10.1103/PhysRevLett.99.221301},
	eprint = {astro-ph/0612294},
	journal = {Phys. Rev. Lett.},
	pages = {221301},
	title = {{Gravitational Wave Production At The End Of Inflation}},
	volume = {99},
	year = {2007}}

@article{Easther:2006gt,
	archiveprefix = {arXiv},
	author = {Easther, Richard and Lim, Eugene A.},
	doi = {10.1088/1475-7516/2006/04/010},
	eprint = {astro-ph/0601617},
	journal = {JCAP},
	pages = {010},
	title = {{Stochastic gravitational wave production after inflation}},
	volume = {04},
	year = {2006}}

@article{Garcia-Bellido:2007nns,
	archiveprefix = {arXiv},
	author = {Garcia-Bellido, Juan and Figueroa, Daniel G.},
	doi = {10.1103/PhysRevLett.98.061302},
	eprint = {astro-ph/0701014},
	journal = {Phys. Rev. Lett.},
	pages = {061302},
	reportnumber = {IFT-UAM-CSIC-06-46},
	title = {{A stochastic background of gravitational waves from hybrid preheating}},
	volume = {98},
	year = {2007}}

@article{Garcia-Bellido:2007fiu,
	archiveprefix = {arXiv},
	author = {Garcia-Bellido, Juan and Figueroa, Daniel G. and Sastre, Alfonso},
	doi = {10.1103/PhysRevD.77.043517},
	eprint = {0707.0839},
	journal = {Phys. Rev. D},
	pages = {043517},
	primaryclass = {hep-ph},
	reportnumber = {IFT-UAM-CSIC-07-38},
	title = {{A Gravitational Wave Background from Reheating after Hybrid Inflation}},
	volume = {77},
	year = {2008}}

@article{Dufaux:2007pt,
	archiveprefix = {arXiv},
	author = {Dufaux, Jean Francois and Bergman, Amanda and Felder, Gary N. and Kofman, Lev and Uzan, Jean-Philippe},
	doi = {10.1103/PhysRevD.76.123517},
	eprint = {0707.0875},
	journal = {Phys. Rev. D},
	pages = {123517},
	primaryclass = {astro-ph},
	title = {{Theory and Numerics of Gravitational Waves from Preheating after Inflation}},
	volume = {76},
	year = {2007}}

@article{Dufaux:2008dn,
	archiveprefix = {arXiv},
	author = {Dufaux, Jean-Francois and Felder, Gary and Kofman, Lev and Navros, Olga},
	doi = {10.1088/1475-7516/2009/03/001},
	eprint = {0812.2917},
	journal = {JCAP},
	pages = {001},
	primaryclass = {astro-ph},
	reportnumber = {FTUAM-08-25, IFT-UAM-CSIC-08-90},
	title = {{Gravity Waves from Tachyonic Preheating after Hybrid Inflation}},
	volume = {03},
	year = {2009}}

@article{Ghiglieri:2015nfa,
    author = "Ghiglieri, J. and Laine, M.",
    title = "{Gravitational wave background from Standard Model physics: Qualitative features}",
    eprint = "1504.02569",
    archivePrefix = "arXiv",
    primaryClass = "hep-ph",
    doi = "10.1088/1475-7516/2015/07/022",
    journal = "JCAP",
    volume = "07",
    pages = "022",
    year = "2015"
}

@article{Ghiglieri:2020mhm,
    author = "Ghiglieri, J. and Jackson, G. and Laine, M. and Zhu, Y.",
    title = "{Gravitational wave background from Standard Model physics: Complete leading order}",
    eprint = "2004.11392",
    archivePrefix = "arXiv",
    primaryClass = "hep-ph",
    doi = "10.1007/JHEP07(2020)092",
    journal = "JHEP",
    volume = "07",
    pages = "092",
    year = "2020"
}

@article{Ringwald:2020ist,
    author = {Ringwald, Andreas and Sch{\"u}tte-Engel, Jan and Tamarit, Carlos},
    title = "{Gravitational Waves as a Big Bang Thermometer}",
    eprint = "2011.04731",
    archivePrefix = "arXiv",
    primaryClass = "hep-ph",
    reportNumber = "DESY 20-187, DESY-20-187, TUM-HEP-1293-20",
    doi = "10.1088/1475-7516/2021/03/054",
    journal = "JCAP",
    volume = "03",
    pages = "054",
    year = "2021"
}

@article{Ghiglieri:2022rfp,
    author = {Ghiglieri, Jacopo and Sch{\"u}tte-Engel, Jan and Speranza, Enrico},
    title = "{Freezing-in gravitational waves}",
    eprint = "2211.16513",
    archivePrefix = "arXiv",
    primaryClass = "hep-ph",
    doi = "10.1103/PhysRevD.109.023538",
    journal = "Phys. Rev. D",
    volume = "109",
    number = "2",
    pages = "023538",
    year = "2024"
}

@article{Ema:2015dka,
    author = "Ema, Yohei and Jinno, Ryusuke and Mukaida, Kyohei and Nakayama, Kazunori",
    title = "{Gravitational Effects on Inflaton Decay}",
    eprint = "1502.02475",
    archivePrefix = "arXiv",
    primaryClass = "hep-ph",
    reportNumber = "UT-15-03",
    doi = "10.1088/1475-7516/2015/05/038",
    journal = "JCAP",
    volume = "05",
    pages = "038",
    year = "2015"
}

@article{Ema:2016hlw,
    author = "Ema, Yohei and Jinno, Ryusuke and Mukaida, Kyohei and Nakayama, Kazunori",
    title = "{Gravitational particle production in oscillating backgrounds and its cosmological implications}",
    eprint = "1604.08898",
    archivePrefix = "arXiv",
    primaryClass = "hep-ph",
    reportNumber = "UT-16-20, KEK-TH-1898, IPMU-16-0062",
    doi = "10.1103/PhysRevD.94.063517",
    journal = "Phys. Rev. D",
    volume = "94",
    number = "6",
    pages = "063517",
    year = "2016"
}

@article{Ema:2020ggo,
    author = "Ema, Yohei and Jinno, Ryusuke and Nakayama, Kazunori",
    title = "{High-frequency Graviton from Inflaton Oscillation}",
    eprint = "2006.09972",
    archivePrefix = "arXiv",
    primaryClass = "astro-ph.CO",
    reportNumber = "DESY-20-104",
    doi = "10.1088/1475-7516/2020/09/015",
    journal = "JCAP",
    volume = "09",
    pages = "015",
    year = "2020"
}

@article{Choi:2024ilx,
    author = "Choi, Gongjun and Ke, Wenqi and Olive, Keith A.",
    title = "{Minimal production of prompt gravitational waves during reheating}",
    eprint = "2402.04310",
    archivePrefix = "arXiv",
    primaryClass = "hep-ph",
    doi = "10.1103/PhysRevD.109.083516",
    journal = "Phys. Rev. D",
    volume = "109",
    pages = "083516",
    year = "2024"
}

@article{Mudrunka:2026kgm,
    author = "Mudrunka, Kamil and Nakayama, Kazunori",
    title = "{High frequency spectrum of primordial gravitational waves}",
    eprint = "2601.00378",
    archivePrefix = "arXiv",
    primaryClass = "hep-ph",
    reportNumber = "TU-1291, KEK-QUP-2025-0028",
    doi = "10.1088/1475-7516/2026/06/050",
    journal = "JCAP",
    volume = "06",
    pages = "050",
    year = "2026"
}

@article{Wang:2026ule,
    author = "Wang, Chenhuan and Xu, Yong and Zhao, Wenbin",
    title = "{Graviton Production from Inflaton Condensate: Boltzmann vs Bogoliubov}",
    eprint = "2604.12687",
    archivePrefix = "arXiv",
    primaryClass = "hep-ph",
    month = "4",
    year = "2026"
}

@article{Wang:2026pff,
    author = "Wang, Yubing and Wu, Quan-feng and Xu, Xun-Jie",
    title = "{A Unified Bogoliubov Approach to Primordial Gravitational Waves: From Inflation to Reheating}",
    eprint = "2604.17478",
    archivePrefix = "arXiv",
    primaryClass = "hep-ph",
    month = "4",
    year = "2026"
}

@article{Xu:2024fjl,
    author = "Xu, Yong",
    title = "{Ultra-high frequency gravitational waves from scattering, Bremsstrahlung and decay during reheating}",
    eprint = "2407.03256",
    archivePrefix = "arXiv",
    primaryClass = "hep-ph",
    reportNumber = "MITP-24-058",
    doi = "10.1007/JHEP10(2024)174",
    journal = "JHEP",
    volume = "10",
    pages = "174",
    year = "2024"
}

@article{Bernal:2025lxp,
    author = "Bernal, Nicol{\'a}s and Wu, Quan-feng and Xu, Xun-Jie and Xu, Yong",
    title = "{Pre-thermalized gravitational waves}",
    eprint = "2503.10756",
    archivePrefix = "arXiv",
    primaryClass = "hep-ph",
    reportNumber = "MITP-25-022",
    doi = "10.1007/JHEP08(2025)125",
    journal = "JHEP",
    volume = "08",
    pages = "125",
    year = "2025"
}

@article{Xu:2025wjq,
    author = "Xu, Xun-Jie and Xu, Yong and Yin, Qiqin and Zhu, Junyu",
    title = "{Full-spectrum analysis of gravitational wave production from inflation to reheating}",
    eprint = "2505.08868",
    archivePrefix = "arXiv",
    primaryClass = "hep-ph",
    reportNumber = "MITP-25-036",
    doi = "10.1007/JHEP10(2025)141",
    journal = "JHEP",
    volume = "10",
    pages = "141",
    year = "2025"
}

@article{Nakayama:2018ptw,
    author = "Nakayama, Kazunori and Tang, Yong",
    title = "{Stochastic Gravitational Waves from Particle Origin}",
    eprint = "1810.04975",
    archivePrefix = "arXiv",
    primaryClass = "hep-ph",
    reportNumber = "UT-18-20",
    doi = "10.1016/j.physletb.2018.11.023",
    journal = "Phys. Lett. B",
    volume = "788",
    pages = "341--346",
    year = "2019",
    note = "[Erratum: Phys.Lett.B 839, 137787 (2023)]"
}

@article{Huang:2019lgd,
    author = "Huang, Da and Yin, Lu",
    title = "{Stochastic Gravitational Waves from Inflaton Decays}",
    eprint = "1905.08510",
    archivePrefix = "arXiv",
    primaryClass = "hep-ph",
    doi = "10.1103/PhysRevD.100.043538",
    journal = "Phys. Rev. D",
    volume = "100",
    number = "4",
    pages = "043538",
    year = "2019"
}

@article{Barman:2023ymn,
    author = "Barman, Basabendu and Bernal, Nicol{\'a}s and Xu, Yong and Zapata, {\'O}scar",
    title = "{Gravitational wave from graviton Bremsstrahlung during reheating}",
    eprint = "2301.11345",
    archivePrefix = "arXiv",
    primaryClass = "hep-ph",
    doi = "10.1088/1475-7516/2023/05/019",
    journal = "JCAP",
    volume = "05",
    pages = "019",
    year = "2023"
}

@article{Barman:2023rpg,
    author = "Barman, Basabendu and Bernal, Nicol{\'a}s and Xu, Yong and Zapata, {\'O}scar",
    title = "{Bremsstrahlung-induced gravitational waves in monomial potentials during reheating}",
    eprint = "2305.16388",
    archivePrefix = "arXiv",
    primaryClass = "hep-ph",
    doi = "10.1103/PhysRevD.108.083524",
    journal = "Phys. Rev. D",
    volume = "108",
    number = "8",
    pages = "083524",
    year = "2023"
}

@article{Hu:2024awd,
    author = "Hu, Weiyu and Nakayama, Kazunori and Takhistov, Volodymyr and Tang, Yong",
    title = "{Gravitational wave probe of Planck-scale physics after inflation}",
    eprint = "2403.13882",
    archivePrefix = "arXiv",
    primaryClass = "hep-ph",
    reportNumber = "KEK-QUP-2024-0006, TU-1226, KEK-TH-2607, KEK-Cosmo-0341, IPMU24-0007",
    doi = "10.1016/j.physletb.2024.138958",
    journal = "Phys. Lett. B",
    volume = "856",
    pages = "138958",
    year = "2024"
}

@article{Hu:2024bha,
    author = "Hu, Wei-Yu and Nakayama, Kazunori and Takhistov, Volodymyr and Tang, Yong",
    title = "{Dual gravitational wave signatures of instant preheating}",
    eprint = "2409.06483",
    archivePrefix = "arXiv",
    primaryClass = "astro-ph.CO",
    reportNumber = "TU-1240, KEK-QUP-2024-0020, KEK-TH-2650, KEK-Cosmo-0356",
    doi = "10.1088/1475-7516/2025/01/029",
    journal = "JCAP",
    volume = "01",
    pages = "029",
    year = "2025"
}

@article{Lee:2026fpg,
    author = "Lee, AlexKen and Mescia, Federico and Wu, Keyun",
    title = "{Enhanced Stochastic Gravitational Waves signals from Wess-Zumino chiral superfield}",
    eprint = "2601.22421",
    archivePrefix = "arXiv",
    primaryClass = "hep-ph",
    month = "1",
    year = "2026"
}

@article{Ananda:2006af,
    author = "Ananda, Kishore N. and Clarkson, Chris and Wands, David",
    title = "{The Cosmological gravitational wave background from primordial density perturbations}",
    eprint = "gr-qc/0612013",
    archivePrefix = "arXiv",
    doi = "10.1103/PhysRevD.75.123518",
    journal = "Phys. Rev. D",
    volume = "75",
    pages = "123518",
    year = "2007"
}

@article{Baumann:2007zm,
    author = "Baumann, Daniel and Steinhardt, Paul J. and Takahashi, Keitaro and Ichiki, Kiyotomo",
    title = "{Gravitational Wave Spectrum Induced by Primordial Scalar Perturbations}",
    eprint = "hep-th/0703290",
    archivePrefix = "arXiv",
    doi = "10.1103/PhysRevD.76.084019",
    journal = "Phys. Rev. D",
    volume = "76",
    pages = "084019",
    year = "2007"
}

@article{Assadullahi:2009nf,
    author = "Assadullahi, Hooshyar and Wands, David",
    title = "{Gravitational waves from an early matter era}",
    eprint = "0901.0989",
    archivePrefix = "arXiv",
    primaryClass = "astro-ph.CO",
    doi = "10.1103/PhysRevD.79.083511",
    journal = "Phys. Rev. D",
    volume = "79",
    pages = "083511",
    year = "2009"
}

@article{Kohri:2018awv,
    author = "Kohri, Kazunori and Terada, Takahiro",
    title = "{Semianalytic calculation of gravitational wave spectrum nonlinearly induced from primordial curvature perturbations}",
    eprint = "1804.08577",
    archivePrefix = "arXiv",
    primaryClass = "gr-qc",
    reportNumber = "KEK-TH-2046, KEK-COSMO-223",
    doi = "10.1103/PhysRevD.97.123532",
    journal = "Phys. Rev. D",
    volume = "97",
    number = "12",
    pages = "123532",
    year = "2018"
}

@article{Inomata:2019zqy,
    author = "Inomata, Keisuke and Kohri, Kazunori and Nakama, Tomohiro and Terada, Takahiro",
    title = "{Gravitational Waves Induced by Scalar Perturbations during a Gradual Transition from an Early Matter Era to the Radiation Era}",
    eprint = "1904.12878",
    archivePrefix = "arXiv",
    primaryClass = "astro-ph.CO",
    reportNumber = "IPMU 19-0066, KEK-TH-2121, KEK-Cosmo-236",
    doi = "10.1088/1475-7516/2019/10/071",
    journal = "JCAP",
    volume = "10",
    pages = "071",
    year = "2019",
    note = "[Erratum: JCAP 08, E01 (2023)]"
}

@article{Nakayama:2008ip,
    author = "Nakayama, Kazunori and Saito, Shun and Suwa, Yudai and Yokoyama, Jun'ichi",
    title = "{Space laser interferometers can determine the thermal history of the early Universe}",
    eprint = "0802.2452",
    archivePrefix = "arXiv",
    primaryClass = "hep-ph",
    reportNumber = "RESCEU-2-08, UTAP-594",
    doi = "10.1103/PhysRevD.77.124001",
    journal = "Phys. Rev. D",
    volume = "77",
    pages = "124001",
    year = "2008"
}

@article{Nakayama:2008wy,
    author = "Nakayama, Kazunori and Saito, Shun and Suwa, Yudai and Yokoyama, Jun'ichi",
    title = "{Probing reheating temperature of the universe with gravitational wave background}",
    eprint = "0804.1827",
    archivePrefix = "arXiv",
    primaryClass = "astro-ph",
    reportNumber = "RESCEU-7-08, UTAP-596",
    doi = "10.1088/1475-7516/2008/06/020",
    journal = "JCAP",
    volume = "06",
    pages = "020",
    year = "2008"
}

@article{Janssen:2014dka,
    author = "Janssen, Gemma and others",
    editor = "Bourke, Tyler L. and others",
    title = "{Gravitational wave astronomy with the SKA}",
    eprint = "1501.00127",
    archivePrefix = "arXiv",
    primaryClass = "astro-ph.IM",
    doi = "10.22323/1.215.0037",
    journal = "PoS",
    volume = "AASKA14",
    pages = "037",
    year = "2015"
}

@article{Bartolo:2016ami,
    author = "Bartolo, Nicola and others",
    title = "{Science with the space-based interferometer LISA. IV: Probing inflation with gravitational waves}",
    eprint = "1610.06481",
    archivePrefix = "arXiv",
    primaryClass = "astro-ph.CO",
    reportNumber = "ACFI-T16-19, UMN-TH-3608-16, CERN-TH-2016-222, KCL-PH-TH-2016-58, IFT-UAM-CSIC-16-104",
    doi = "10.1088/1475-7516/2016/12/026",
    journal = "JCAP",
    volume = "12",
    pages = "026",
    year = "2016"
}

@article{Yagi:2011wg,
    author = "Yagi, Kent and Seto, Naoki",
    title = "{Detector configuration of DECIGO/BBO and identification of cosmological neutron-star binaries}",
    eprint = "1101.3940",
    archivePrefix = "arXiv",
    primaryClass = "astro-ph.CO",
    doi = "10.1103/PhysRevD.83.044011",
    journal = "Phys. Rev. D",
    volume = "83",
    pages = "044011",
    year = "2011",
    note = "[Erratum: Phys.Rev.D 95, 109901 (2017)]"
}

@article{Punturo:2010zz,
    author = "Punturo, M. and others",
    editor = "Ricci, Fulvio",
    title = "{The Einstein Telescope: A third-generation gravitational wave observatory}",
    doi = "10.1088/0264-9381/27/19/194002",
    journal = "Class. Quant. Grav.",
    volume = "27",
    pages = "194002",
    year = "2010"
}

@article{Schmitz:2020syl,
    author = "Schmitz, Kai",
    title = "{New Sensitivity Curves for Gravitational-Wave Signals from Cosmological Phase Transitions}",
    eprint = "2002.04615",
    archivePrefix = "arXiv",
    primaryClass = "hep-ph",
    reportNumber = "CERN-TH-2020-018",
    doi = "10.1007/JHEP01(2021)097",
    journal = "JHEP",
    volume = "01",
    pages = "097",
    year = "2021"
}

@article{Herman:2020wao,
    author = {Herman, Nicolas and F{\"u}zfa, Andr{\'e} and Lehoucq, L{\'e}onard and Clesse, S{\'e}bastien},
    title = "{Detecting planetary-mass primordial black holes with resonant electromagnetic gravitational-wave detectors}",
    eprint = "2012.12189",
    archivePrefix = "arXiv",
    primaryClass = "gr-qc",
    doi = "10.1103/PhysRevD.104.023524",
    journal = "Phys. Rev. D",
    volume = "104",
    number = "2",
    pages = "023524",
    year = "2021"
}

@article{Herman:2022fau,
    author = "Herman, Nicolas and Lehoucq, L{\'e}onard and F{\'{u}}zfa, Andr{\'e}",
    title = "{Electromagnetic antennas for the resonant detection of the stochastic gravitational wave background}",
    eprint = "2203.15668",
    archivePrefix = "arXiv",
    primaryClass = "gr-qc",
    doi = "10.1103/PhysRevD.108.124009",
    journal = "Phys. Rev. D",
    volume = "108",
    number = "12",
    pages = "124009",
    year = "2023"
}

@article{Blas:2026yqb,
    author = {Blas, Diego and Gasparotto, Silvia and Murayama, Hitoshi and Sch{\"u}tte-Engel, Jan and Zosso, Jann},
    title = "{Low-Frequency Gravitational Bremsstrahlung and Memory in a Medium}",
    eprint = "2608.24994",
    archivePrefix = "arXiv",
    primaryClass = "hep-ph",
    reportNumber = "RIKEN-iTHEMS-Report-26",
    month = "8",
    year = "2026"
}

@article{Mukaida:2026bdh,
    author = "Mukaida, Kyohei and Tsuji, Tenta",
    title = "{Gravitational Waves from Reheating beyond Instantaneous Thermalization}",
    eprint = "2609.01480",
    archivePrefix = "arXiv",
    primaryClass = "hep-ph",
    reportNumber = "KEK-TH-2868",
    month = "9",
    year = "2026"
}

@article{Figueroa:2020rrl,
    author = "Figueroa, Daniel G. and Florio, Adrien and Torrenti, Francisco and Valkenburg, Wessel",
    title = "{The art of simulating the early Universe -- Part I: Integration techniques and canonical cases}",
    eprint = "2006.15122",
    archivePrefix = "arXiv",
    primaryClass = "astro-ph.CO",
    doi = "10.1088/1475-7516/2021/04/035",
    journal = "JCAP",
    volume = "04",
    pages = "035",
    year = "2021"
}

\end{document}